\documentclass[fleqn,usenatbib]{mnras}
\usepackage{microtype}

\usepackage{newtxtext}
\usepackage[varg]{newtxmath}

\usepackage[T1]{fontenc}
\usepackage{ae,aecompl}

\usepackage{graphicx}	
\usepackage{amsmath}	
\usepackage{amssymb}	
\usepackage{tikz}

\usepackage{color}

\title[\textit{Thyr}: A 3D Volumetric Microwave Simulator]{\textit{Thyr}: A Volumetric Ray-Marching Tool for Simulating Microwave Emission}

\author[C. M. J. Osborne \& P. J. A. Sim\~{o}es]{
Christopher M. J. Osborne,$^{1}$\thanks{E-mail: c.osborne.1@research.gla.ac.uk}
Paulo J. A. Sim\~{o}es,$^{1,2,3}$
\\
$^{1}$SUPA, School of Physics and Astronomy, University of Glasgow, Glasgow, G12 8QQ, UK\\
$^{2}$Centro de R\'adio Astronomia e Astrof\'isica Mackenzie, CRAAM, Universidade Presbiteriana Mackenzie, S\~ao Paulo, SP 01302-907, Brazil\\
$^{3}$Escola de Engenharia, Universidade Presbiteriana Mackenzie, S\~ao Paulo, SP 01302-907, Brazil\\
}
\date{Accepted XXX. Received YYY; in original form ZZZ}

\pubyear{2018}

\begin{document}
\label{firstpage}
\pagerange{\pageref{firstpage}--\pageref{lastpage}}
\maketitle

\begin{abstract}
Gyrosynchrotron radiation is produced by solar flares, 
and can be used to infer 
properties of the accelerated electrons and magnetic 
field of the flaring region. 
This microwave emission is highly dependent on many 
local plasma parameters, 
and the viewing angle.
To correctly interpret observations, detailed simulations of 
the emission are required.
Additionally, gyrosynchrotron emission from the chromosphere 
has been largely ignored in modelling efforts, and recent studies have 
shown the importance of thermal emission at millimetric wavelengths. \textit{Thyr} is a new tool for modelling microwave
emission from three-dimensional flaring loops with spatially varying
atmosphere and increased resolution in the lower corona and chromosphere.
\textit{Thyr} is modular and open-source, consisting of
separate components to compute the thermal and non-thermal microwave
emission coefficients and perform three-dimensional radiative transfer 
(in local thermodynamic equilibrium). 
The radiative transfer integral is computed by a novel ray-marching
technique to efficiently compute the contribution of 
many volume elements.
This technique can also be employed on a variety of astrophysics problems.
Herein we present a review of the theory of gyrosynchrotron 
radiation, and two simulations of identical flare loops 
in low- and high-resolution performed with \textit{Thyr}, 
with a spectral imaging analysis of differing regions.
The high-resolution simulation presents a spectral hardening at
higher frequencies. 
This hardening originates around the top of the chromosphere due 
to the strong convergence of the magnetic field, 
and is not present in previous models due to insufficient resolution.
This hardening could be observed with a coordinated flare observation from 
active radio observatories.
\end{abstract}

\begin{keywords}
Sun: flares -- Sun: radio radiation -- radiative transfer -- methods: numerical
\end{keywords}

\section{Introduction and Background}

Strong increases in microwave emission are observed during solar flares.
Gyrosynchrotron emission, originating from mildly relativistic
non-thermal electrons spiraling around magnetic field lines, is
responsible for the majority of the emission in the 1-200 GHz range \citep{Dulk1985,Bastian1998}.
Recent studies have also shown the importance of free-free emission from thermal electrons 
at these temperatures \citep[e.g.][]{Heinzel2012}.
Observations of the microwave emission produced during a solar flare can be
used to characterise the properties of the accelerated electrons and magnetic field of the flaring region.

Current observations of solar flares are typically carried out by radio telescopes without imaging 
capabilities (\textit{Sun-as-a-star}) and by the interferometer \textit{Nobeyama Radio Heliograph} 
\citep[NoRH,][]{Nakajima1994}, capable of producing images at 17 and 34 GHz with moderate spatial 
resolution (10$''$ and 5$''$, respectively) and with a temporal resolution down to 1 s. 
Such observations are capable of resolving the radio flare sources, 
associated with footpoints 
\citep[e.g.][]{Tzatzakis2008}, flaring loops \citep[e.g.][]{Kundu2001}, or ribbons \citep[e.g.][]{Simoes2013a}
These features are commonly associated with the geometry of magnetic field loops, filled with 
electrons accelerated during the energy release phase of flares.

The first microwave spectral imaging analysis of a flare, with high spectral 
resolution, was performed by \citet{Wang1994}, using the \textit{Owens Valley Radio Observatory} \citep[OVSA, ][]{Lim1994}
1--18~GHz data, and showed that the microwave spectral peak occurs at lower frequencies ($\nu < 10$~GHz) for 
the looptop sources and at higher frequencies ($\nu > 10$~GHz)in the footpoints, 
following the strength of the magnetic field in a flaring loop. 
Another example of a microwave imaging spectroscopic analysis was presented by \cite{Wang1995}, also using OVSA data.
A similar study was recently performed by \citet{Gary2018} with the
\textit{Extended Owens Valley Solar Array} \citep[EOVSA, ][]{Kuroda2017, Wang2015} observations 
of the SOL2017-09-10 X8.2 event and yielded 
results consistent with those of \citet{Wang1994} and \citet{Wang1995}, and simulation results of \cite{Simoes2006}.
\citet{Nindos2000} used one-dimensional modelling to reproduce the spatially 
resolved emission of a flare loop in 5 and 15 GHz, showing the 15 GHz 
emission was produced only in the footpoints 
and the 5 GHz emission restricted to the upper loop.
To fit the observations it was necessary to use a much stronger magnetic field 
in the photosphere than at the looptop (870~G in the footpoints and 280~G in the looptop).
Analysis of similar events by \citet{Lee2000} provided evidence for magnetic trapping that would 
restrict the 5 GHz emission to the upper loop while the 15 GHz emission comes 
from higher energy electrons that are able to escape mirroring effects.

The complexity of solar microwave simulations has increased significantly in recent years, primarily building on the three-dimensional modeling tool GS Simulator (now GX Simulator) \citep{Kuznetsov2011,Nita2015}. 
GX Simulator computes solar radio emission from a three dimensional model, with in-built tools for magnetic field extrapolation from photospheric magnetograms, and a chromospheric approximation computed using the semi-empirical atmospheres of \citet{Fontenla2009}, based on the photospheric intensity and magnetogram values \citep{Fleishman2015}.
This tool has been used to forward fit and reconcile observations of radio and hard X-ray emission with a unified electron population \citep{Kuroda2017}, and investigate the plasma heating mechanism during a ``cold'' flare event using Linear Force Free Field extrapolation on magnetogram data to reconstruct the observed two-loop geometry and explain the heating delay by electron trapping in the looptop \citep{Fleishman2016}.
\citet{Gordovskyy2017} used GX Simulator to investigate the polarised microwave emission from relaxing twisted coronal loops based on time-dependent magnetohydrodynamic simulations.
These loops were found to produce gyrosynchrotron emission with short-term gradients of circular polarisation (cross-loop polarisation gradients), that depend strongly on viewing angle, and would primarily be visible on a loop observed on the solar limb, clearly showing the three-dimensional nature of the problem.

With the arrival of new and upgraded solar observatories, such as the \textit{Atacama Large Millimetric-submillimetric Array} 
\citep[ALMA, ][]{Wedemeyer2016} and EOVSA
that are providing higher spatial and spectral resolution it is essential to have
detailed predictions and models for these observations.
While solar observations with ALMA have started \citep{White2017,Iwai2017,Brajsa2017}, the Sun is now at its period of minimum activity, and no flares have been detected with ALMA yet. However, ALMA's capability for flare studies have been proven with the detection of a super-flare on Proxima Centauri \citep{MacGregor2018}.

Regular solar observations in the millimetric range have started in 1999 with the \textit{Solar Submillimeter Telescope} \citep[SST, operating at 212 and 405 GHz][]{Kaufmann2008}. Since then a number of solar flares have been detected, with the typical decreasing, non-thermal spectrum towards higher frequencies \citep{GimenezdeCastro2009,GimenezdeCastro2013}, evidence for thermal (free-free) emission (especially at high frequencies towards the infrared range) \citep{Heinzel2012, Simoes2017, Trottet2012, Trottet2015}, and with an increasing spectrum component at millimetric wavelengths (sub-THz) \citep[e.g.][]{Kaufmann2004}, which was also observed with the \textit{K\"oln Observatory for Submillimeter and Millimeters Astronomy} \citep[KOSMA, ][]{Luthi2004}. This \textit{sub-THz component} still remains without a satisfactory physical explanation \citep{Krucker2013}.

Previous observations have shown that flaring ribbons and footpoints can reach temperatures in the range of 1-10~MK
\citep{Hudson1994, Mrozek2004, Graham2013, Simoes2015a}. 
The chromospheric plasma is normally opaque to radio emission (in the GHz range), and it would therefore be irrelevant if non-thermal electrons penetrate this layer. However, if this plasma is heated to these greater temperatures, then the free-free opacity drops and the plamsa becomes optically thin to any gyrosynchrotron emission potentially produced in the chromosphere, as discussed in \citet{Fletcher2013}.

Herein we present a new tool, \textit{Thyr}, to compute the microwave emission from
a dipole loop of plasma, containing a spatially variable atmosphere.
Emission maps are computed under the assumption of local thermodynamic
equilibrium (LTE) in the three-dimensional dipole volume.
Full spectral information is available for each simulated pixel and can also be
integrated over larger regions.

\textit{Thyr's} novel feature is its ability to manually refine the simulation resolution in the chromosphere while maintaining a complete microwave radiation treatment,
allowing it to resolve the much smaller scales over which the atmospheric parameters
evolve helping to account for both the free-free and gyrosynchrotron emission from this thin layer.
Our model also supports using arbitrarily varying atmospheric parameters, magnetic field configurations, and electron pitch angle distributions.

\textit{Thyr} builds on several generations of gyrosynchrotron modelling
tools, including, but not limited to, \citet{Klein1984}, \cite{Alissandrakis1984}, \cite{Holman2003}, \cite{Simoes2006,
Simoes2010}. 
\citet{Costa2013} and \citet{Cuambe2018} have both constructed
large databases of three-dimensional gyrosynchrotron simulations from a 
large parameter space to develop an understanding of solar bursts and attempt 
to infer flare parameters from observation respectively. 
GX Simulator \citep{Nita2015} can produce three-dimensional simulations of 
gyrosynchrotron emission from a dipole loop and has been used to investigate the 
effects of varying electron pitch-angle distribution \citep{Kuznetsov2011}, but focuses primarily on coronal microwave emission.
These tools, including \textit{Thyr}, build on the analytic expressions describing
gyrosynchrotron emission formulated by \citet{Ramaty}.

In this paper we describe the functioning of the \textit{Thyr} tool, validate
its output against simulations presented in \citet{Klein1984}, and 
use new simulations demonstrating the importance of modelling the lower atmosphere at 
high-resolution to capture the fine structures of the chromosphere.
The source code and examples are available on Github at \url{http://github.com/Goobley/Thyr2} \citep{Osborne2018}.


\section{Theory}

\subsection{Gyrosynchrotron Emission}\label{Sec:GS}

Gyrosynchrotron emission depends on a large number of parameters of the emitting region. These are primarily \citep{Ramaty,Sthli1989}

\begin{itemize}
\item Magnetic field strength $B$: Directly determines the electron gyrofrequency.
\item Plasma density $n_p$: Describes the density of electrons in the thermal plasma and can have a large effect on
  measured emission due to free-free emission and absorption effects. Dense magnetised plasmas may also strongly suppress the 
  gyrosynchrotron emission, an effect known as
  Razin suppression \citep{Ramaty}.
\item Non-thermal electron density $n_e$: This parameter is largely responsible for the strength of the
  emission, as a higher electron density will produce more radiation. It also affects the importance of the gyrosynchrotron self-absorption, via radiative transfer.
\item Non-thermal electron distribution $g(\gamma, \phi)$: This function describes the distribution of non-thermal 
  electron energies (here in terms of relativistic Lorentz factor $\gamma$) and pitch-angles -- the angle between the electron's 
  velocity vector and the magnetic field. It is often assumed that the electron energies follow a single power law distribution
  determined by their spectral index $\delta$, $f(\gamma) \propto (\gamma-1)^{-\delta} d\gamma$, and minimum and maximum 
  cut-off energy values, $\gamma_\mathrm{min}$ and $\gamma_\mathrm{max}$, respectively.
  The distribution of pitch-angle $\phi$ has a strong effect on the angles radiation is observed at due to the strong beaming
  effect of the radiation from relativistic electrons.
  \textit{Thyr} supports multi-power law energy distributions and arbitrary distributions of pitch-angle.
\item Viewing angle $\theta$: The angle between the wave vector and the magnetic line has a strong effect on
  the observed radiation due to the polarisation of the radiation and the beaming of emission from relativistic particles.
\end{itemize}

As gyrosynchrotron emission is produced by electrons spiraling around the magnetic 
field, it is composed of two circularly polarised modes.
We designate the mode extraordinary when the circular polarisation is right-handed 
and the electric field vector rotates in the same direction as the electrons.
In the opposite case the mode is called ordinary, or left-handed.
Following the convention of \citet{Klein1984} we use ``$-$'' to refer to the 
extraordinary mode (right-handed polarisation for positive $B$) in 
equations, and ``$+$'' to refer to the ordinary mode (left-handed).

The following treatment is based on that of \citet{Klein1984}, 
developed from \citet{Ramaty} and the corrections of \citet{Trulsen1970}.
The refractive index of the plasma is given by the Appleton-Hartree equation (e.g. \citet{Stix1962})

\begin{equation}
\begin{split}
&n _ { \pm }(\nu, \theta) ^ { 2 } = 1 - \\ &\frac { 2 X ( 1 - X ) } { 2 ( 1 - X ) - Y ^ { 2 } \sin ^ { 2 } \theta \pm [ Y ^ { 4 } \sin ^ { 4 } \theta + 4 Y ^ { 2 } ( 1 - X ) ^ { 2 } \cos ^ { 2 } \theta ] ^ { 1 / 2 } }
\end{split}
\end{equation}

where
\begin{equation*}
\begin{split}
X&:=\frac{\nu_p^2}{\nu^2},\\
Y&:=\frac{\nu_B}{\nu},\\
\nu_p &:= \textrm{electron plasma frequency} = e\sqrt{\frac{n_p}{\pi m_e}},\\
\nu_B &:= \textrm{electron gyrofrequency} = \frac{eB}{2\pi m_e c},\\
e &:= \textrm{electron charge,}\\
m_e &:= \textrm{electron mass,}\\
\end{split}
\end{equation*}

The polarisation coefficient $a_\theta$ is then

\begin{equation}
a _ { \theta \pm }(\nu,\theta) = \frac { 2 ( 1 - X ) \cos \theta } { - Y \sin ^ { 2 } \theta \pm [ Y ^ { 2 } \sin ^ { 4 } \theta + 4 ( 1 - X ) ^ { 2 } \cos ^ { 2 } \theta ] ^ { 1 / 2 } }
\end{equation}

and is used to compute the spectral intensity $\epsilon$ of a single electron \citep{Trulsen1970}

\begin{equation}
\begin{split}
\epsilon _ { \pm }(\nu, \theta, \gamma, \phi) &= \frac { 2 \pi e ^ { 2 } } { c } \nu ^ { 2 } \frac { n _ { \pm } ^ { 2 } } { 1 + a _ { \theta \pm } ^ { 2 } }\\ &\cdot \sum _ { s = - \infty } ^ { \infty } \left[ - \beta \sin \phi J _ { s } ^ { \prime } ( x _ { s } ) + a _ { \theta \pm } \frac { \cos \theta - n _ { \pm } \beta \cos \phi } { n _ { \pm } \sin \theta } J _ { s } ( x _ { s } ) \right] ^ { 2 } \\ &\cdot \delta [ (1 - n_\pm\beta\cos{\phi}\cos{\theta})\nu - s\nu_B/\gamma ]
\end{split}
\end{equation}

where

\begin{equation*}
\begin{split}
c &:= \textrm{speed of light,}\\
\gamma &:= (1-\beta^2)^{-1/2},\\
\beta &:= \frac{v}{c},\\
x_s &:= \frac{\gamma\nu}{\nu_B} n_\pm\beta \sin{\phi} \sin{\theta},\\
J_s &:= \textrm{Bessel function of the first kind, of order s,} \\&\hspace{3em}\textrm{and }J_s^\prime\textrm{ its derivative with respect to }x_s,\\
\delta &:= \textrm{Dirac delta function}.
\end{split}
\end{equation*}

The electron distribution is described by the function $g(\gamma, \phi)$ such that

\begin{equation}
2 \pi \int _ { 1 } ^ { \infty } d \gamma \int _ { - 1 } ^ { 1 } d ( \cos \phi ) g ( \gamma , \phi ) = 1.
\end{equation}

This gives rise to the expression of the emission and absorption coefficients, $j$ and $\kappa$ respectively,
for a homogeneous ensemble of electrons with density $n_{e}$ and energy and pitch-angle distribution $g$

\begin{equation}
\label{eq:AnaJ}
j _ { \pm } ( \nu , \theta ) = 2 \pi n_{e} \int _ { 1 }^\infty d \gamma \int _ { - 1 }^1 d ( \cos \phi ) g ( \gamma , \phi ) \epsilon_ { \pm } ( \nu , \theta , \gamma , \phi ),
\end{equation}

\begin{equation}
\begin{split}
\label{eq:AnaK}
\kappa _ { \pm } ( \nu , \theta ) = \frac { 2 \pi n_{e}} { m_e \nu ^ { 2 } n _ { \pm } } \int_1^\infty d \gamma \int _ { - 1 }^1 d ( \cos \phi ) \epsilon _ { \pm } ( \nu, \theta , \gamma , \phi )h_\pm(\nu, \theta, \gamma, \phi)
\end{split}
\end{equation}

where 

\begin{equation}
h_\pm(\nu, \theta, \gamma, \phi) = \left[ - \beta \gamma ^ { 2 } \frac { \partial } { \partial \gamma } \left( \frac { g ( \gamma , \phi ) } { \beta \gamma ^ { 2 } } \right) + \frac { n _ { \pm } \beta \cos \theta - \cos \phi } { \beta ^ { 2 } \gamma \sin \phi } \frac { \partial } { \partial \phi } g ( \gamma , \phi )\right].
\end{equation}

The presence of the delta function in $\epsilon$ allows the integral over $\cos{\phi}$ to be solved analytically for $\theta \neq \pi/2$.

Following \citet{Ramaty} we define $G_\pm$ and $H_\pm$

\begin{equation}
\begin{split}
\label{Eq:G}
G_\pm(\nu, \theta, \gamma, s) =&\left[-\beta\sin{\phi_s} J_s^\prime(x_s) + a_{\theta\pm}\left(\frac{\cos{\theta}}{n_\pm \sin{\theta}} - \beta\frac{\cos{\phi_s}}{\sin{\theta}}\right)J_s(x_s)\right]^2\\
&\cdot\frac{g(\gamma, \phi)}{\beta(1+a_{\theta\pm}^2)}\frac{2\pi}{\cos{\theta}}\frac{\nu}{\nu_B},
\end{split}
\end{equation}

\begin{equation}
\begin{split}
\label{Eq:H}
H_\pm(\nu, \theta, \gamma, s) = G_\pm(\nu, \theta, \gamma, s) \cdot h_\pm(\nu, \theta, \gamma, \phi_s)
\end{split}
\end{equation}

where

\begin{equation*}
\begin{split}
\cos{\phi_s} &:= \frac{1-\frac{s\nu_B}{\gamma\nu}}{n_\pm\beta\cos{\theta}},\\
x_s &:= \frac{sn_\pm\beta\sin{\theta}\sin{\phi_s}}{1 - n\pm\beta\cos{\theta}\cos{\phi_s}}.
\end{split}
\end{equation*}

Solving the integral over $\cos{\phi}$ analytically and swapping the 
resultant integral over $\gamma$ with the summation over $s$ by following 
the approach of \citet{Holman2003} to equations \eqref{eq:AnaJ} and \eqref{eq:AnaK} yields

\begin{equation}
j_\pm(\nu,\theta) = \frac{e^3Bn_e}{m_ec^2}\sum_{s=s_\textrm{min}}^{s_\textrm{max}}\int_{\gamma_\textrm{min}}^{\gamma_\textrm{max}} d\gamma G_\pm(\nu, \theta, \gamma, s),
\end{equation}

\begin{equation}
\kappa_\pm(\nu,\theta) = \frac{4\pi e^2n_e}{B}\sum_{s=s_\textrm{min}}^{s_\textrm{max}}\int_{\gamma_\textrm{min}}^{\gamma_\textrm{max}} d\gamma H_\pm(\nu, \theta, \gamma, s),
\end{equation}

where

\begin{equation*}
\begin{split}
s_\textrm{min} &:= \left\lfloor \frac{\gamma\nu}{\nu_B}(1-n_\pm\beta\cos{\theta}) \right\rfloor + 1,\\
s_\textrm{max} &:= \left\lfloor \frac{\gamma\nu}{\nu_B}(1+n_\pm\beta\cos{\theta}) \right\rfloor,\\
\gamma_{\textrm{min,max}} &:= \frac{\frac{s\nu_B}{\nu} \pm n_\pm\cos{\theta}\left[\left(\frac{s\nu_B}{\nu}\right)^2 + n_\pm^2\cos^2{\theta}-1\right]^{1/2}}{1-n_\pm^2\cos^2{\theta}},
\end{split}
\end{equation*}

and $\lfloor x \rfloor$ is a function that returns the floor of $x$.

In reality, we do not always compute the summation up to 
$s_\textrm{max}$ as $G_\pm$ and $H_\pm$ may have converged for a smaller $s$.
This is determined by the relative change of $G_\pm$ and $H_\pm$ across successive 
iterations of the summation.

Additionally, ordinary-mode emission may only be produced when $\nu>\nu_p$ and $n_+>0$.
Similarly, extraordinary-mode emission requires
\begin{equation}
\label{Eq:ExtraProp}
\nu > \left(\nu_p^2 + \frac{1}{4}\nu_B^2\right)^{1/2}+\frac{1}{2}\nu_B
\end{equation}
and $n_- > 0$.

In our implementation the integrals over $\gamma$ are computed using a
Gauss-Legendre method. 
Additionally, significant speed-ups were obtained 
by using a look-up table for the common range of Bessel functions
encountered during the simulation.
This is very efficient because the computation of the Bessel function is by
far the dominant computational cost in the gyrosynchrotron coefficients and
similar regions of the Bessel function will be accessed sequentially, allowing
the CPU's caching and pre-fetching mechanisms to work efficiently.
If a Bessel function outside the tabled space is requested, and 
the approximation is valid, then the expression from \citet{Wild1971} is used.

The expressions for $G_\pm$ \eqref{Eq:G} and $H_\pm$ \eqref{Eq:G} do not hold for 
$\theta=90^\circ$, whilst it is possible to derive additional expressions for this case, there is little reason to, as it only holds for this singular case, and also for reasons discussed in Sec.~\ref{Sec:RT}.

\subsection{Thermal Emission}

In addition to the radio gyrosynchrotron emission parameters, we compute the
thermal free-free emission and absorption coefficients, as well as the \textsc{H$^-$} opacity, known to be relevant for submillimetric emission \citep{Heinzel2012,Simoes2017}.

For the free-free opacity we follow \citet{Dulk1985}, with the correction from \citet{Wedemeyer2016}

\begin{equation}
\begin{aligned}
\kappa_{\nu,\textrm{ff}} &= 9.78\times10^{-3}\frac{n_e}{\nu^2T^{3/2}}\sum_iZ_i^2n_i\\
&\times\begin{cases}
(17.9 + \ln{T^{3/2}} - \ln{\nu}),\quad (T < 2\times10^5\textrm{K})\\
(24.5 + \ln{T} - \ln{\nu}),\quad (T > 2\times10^5\textrm{K})
\end{cases}
\end{aligned}
\end{equation}

where $n_e$ is the thermal electron number density, $n_i$ is the number density of
ion $i$ with charge $Z_i$, $\nu$ is the frequency, and $T$ is the temperature
of the plasma. 
Herein we assume a uniform hydrogen plasma such that

\begin{equation}
\sum_iZ_in_i = n_p = n_{\textsc{Hii}}
\end{equation}

where $n_p$ is the proton density and $n_\textsc{Hii}$ is the ionised hydrogen density.

Now, by \citet{Dulk1985}  (and Kirchoff's law) we have the free-free emission coefficient

\begin{equation}
j_{\nu,\textrm{ff}} = \frac{2k_BT\nu^2k_{\nu,\textrm{ff}}}{c^2}
\end{equation}

where $k_B$ is Boltzmann's constant.

We follow the treatment of \citet{Kurucz1970} computing the \textsc{H$^-$} opacity

\begin{equation}
\kappa_{\nu,\textsc{H$^-$}} = \frac{n_en_\textsc{H}}{\nu}(A_1 + (A_2 - A_3 / T) / \nu) (1 - e^{-h\nu/k_BT})
\end{equation}

where $n_\textsc{H}$ is the neutral hydrogen density, $A_1~=~1.3727~\times~10^{-25}$,
$A_2~=~4.3748~\times~10^{-10}$, and $A_3~=~2.5993~\times~10^{-7}$.

Finally, we have the thermal emission and absorption coefficients

\begin{equation}
\begin{aligned}
j_{\nu,\textrm{therm}} &= j_{\nu,\textrm{ff}} + \kappa_{\nu,\textsc{H-}}\,B_\nu(T),\\
\kappa_{\nu,\textrm{therm}} &= \kappa_{\nu,\textrm{ff}} + \kappa_{\nu,\textsc{H-}},
\end{aligned}
\end{equation}

where $B_\nu(T)$ is the Planck function.
At solar gyrosynchrotron frequencies free-free opacity always dominates over H{\textsc{$-$}} emissivity, and therefore the effects of H{\textsc{$-$}} emissivity were ignored during the numerical simulations.

\begin{figure*}
   \centering
   \includegraphics[width=0.25\textwidth,trim=10 30 0 0,clip]{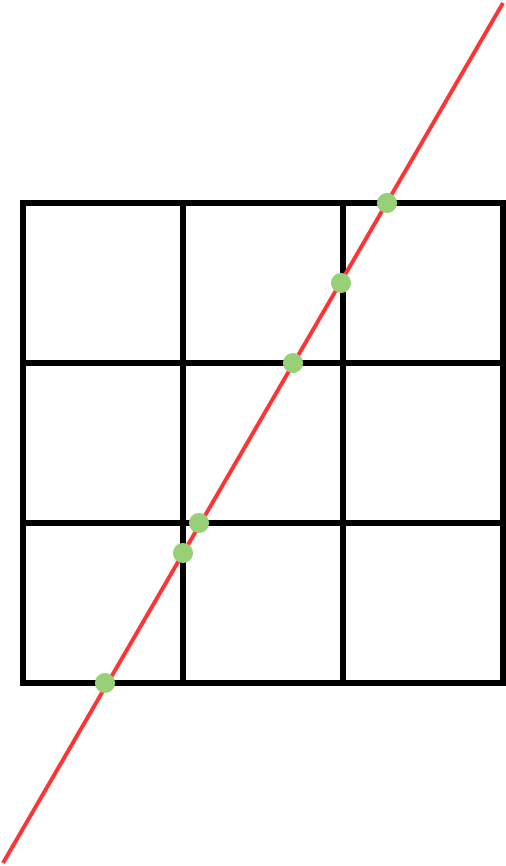}
   \hspace{2cm}
   \includegraphics[width=0.25\textwidth,trim=0 0 0 0,clip]{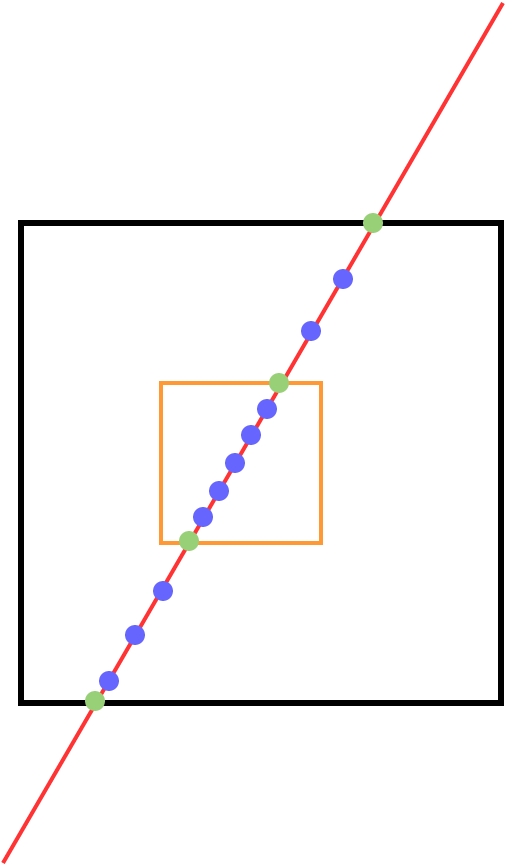}
      \caption{\textit{Left:} Schematic diagram of the process involved in computing the radiative transfer integral through ray-tracing. Here the ray is shown in red, and the intersections with the voxel boundaries in green. \textit{Right:} Schematic diagram of the process involved in computing the radiative transfer integral through ray-marching. Here the ray is shown in red, the intersections that need to be computed in green, the points where the volume texture is sampled in blue, and a region of increased resolution in orange. 
      }
         \label{Fig:RT}
\end{figure*}

\subsection{Radiative Transfer}\label{Sec:RT}

To compute the emission maps generated by this process we compute radiative transfer 
under the assumption of local thermodynamic equilibrium.
The monochromatic observed intensity $I_\nu$ is then given by the radiative transfer equation 
\begin{equation}
    \frac{dI_{\nu}}{d\tau_\nu}=-I_\nu(\tau_\nu)+S_\nu(\tau_\nu),
\end{equation}

where $\tau_\nu$ is the monochromatic optical length of the plasma at frequency $\nu$ along the photon's path between emission and absorption, and $S_\nu$ is the monochromatic source function where $S_\nu = j_\nu / \kappa_\nu$.

Using the ray-marching algorithm of Section \ref{Sec:RayMarch} these emission and
absorption parameters are combined, from back to front 
along the observer's line of sight,
so as to produce a brightness temperature map for each radio mode and
the thermal emission ($T_{b,+}$, $T_{b,-}$, and $T_{b,\textrm{therm}}$
respectively).
The brightness temperature is computed for each pixel of each map as 
per the Rayleigh-Jeans approximation

\begin{equation}
  T_{b} = \frac{c^2}{2k_B\nu^2}I_{\nu}\\
\end{equation}

where

\begin{align}
  I_{\nu} &= \int_{0}^H j_{\nu,o} e^{-\tau_{\nu}}\,dh,\\
  \tau_{\nu} &= \int_{0}^H k_{\nu}\,dh
\end{align}

 and the optical length travelled by a photon $\tau_\nu$ is related to the
 depth $h$ along the observer's line of sight by $d\tau_\nu = -k_\nu\,dh$, 
 where $k_\nu$ is the sum of the absorption coefficients from different processes 
 affecting the frequency $\nu$.
 In these integrals, $H$ is the photon emission point along the observer's line of sight, and 0 is the observer's location.

 A total emission map is then computed from $T_{b,+} + T_{b,-} + T_{b,\textrm{therm}}$.
 
\subsection{Circular Polarisation Degree} 
 
When computing the total emission it is sufficient to simply add the flux across 
the emitting modes, however, for investigating polarisation it is necessary to 
include propagation effects and the radiative transfer of all four Stokes parameters.
Following \citet{Simoes2006} and \citet{Cohen1960}, from magnetoionic theory in solar 
conditions there are two regimes for the propagation of circular polarisation (Stokes V), 
the quasi-longitudinal and quasi-transversal approximations.

Using the C7 semi-empirical atmosphere \citep{Avrett2008} and the dipole model discussed
in Sec.~\ref{Sec:Dipole} with a looptop magnetic field strength of 100 G the quasi-longitudinal 
approximation holds for $\theta$ outside the range $(87^\circ-93^\circ)$ at $\tau=1$ optical 
depth from free-free absorption.
For cells where $\theta$ falls within the $(87^\circ-93^\circ)$ range we simply adjust the viewing angle 
to the closer edge of the range to retain the validity of the quasi-longitudinal approximation.
Unless a very highly anisotropic pitch angle distribution is used, that is peaked within this 
range, there are no significant differences in the intensities of the two modes due to the 
application of this approximation.

From the optical depth calculation we find that 10 GHz radiation does not penetrate 
through the transition region, 45 GHz radiation just enters the chromosphere, and 200 
GHz radiation reaches $\sim$400 km into the chromosphere.
In regions deeper in the chromosphere, for viewing angles close to perpendicular, 
the quasi-longitudinal approximation may fail, and one should be wary of interpreting 
the polarisation results from this region. 
The calculation of the total intensity is, however, unaffected by the use of this approximation.

Following the creation of emission maps we can compute the circular polarisation
degree of the radiation.
We follow the convention of \citet{Klein1984} and define for the Stokes $V$ and $I$ in the 
frame of the wave

\begin{equation}
  \frac{V}{I} = \textrm{sign}(\cos\theta)\,\frac{T_{b,+} - T_{b,-}}{T_{b,+} + T_{b,-} + T_{b,\textrm{therm}}}.
\end{equation}

In the reference frame of an observer the circular polarisation will be reversed such that $V_\textrm{obs}=-V_\textrm{wave}$.

In \emph{Thyr} the magnetic viewing angle term used for calculating the polarisation degree is simply computed as an average along the ray, and so caution should be taken with the interpretation of polarised radiation from overlapping sources with opposing magnetic field orientation. The viewing angles discussed in the previous sections are all treated without this approximation.

\section{The \emph{Thyr} Model}
\subsection{Magnetic field geometry}\label{Sec:Dipole}

We adopt the same analytic dipole model as \citet[see Appendix]{Kuznetsov2011} 
to describe the magnetic field geometry. 
This describes both the magnetic field at each point in the volume and the
region contained within the dipole. 
This analytic dipole model is described in terms of the magnetic field at
the centre of the looptop, the radius of the flux tube at the looptop, the height, 
and the submerged depth of the dipole. 
The magnetic dipole element at the base of the loop can also be inclined, and
the ray-marching method allows the loop to be visualised from any angle which
can be specified in terms of location on the Sun. A solar location is specified
by four angles, solar latitude and longitude, tilt away from the local normal
to the surface about an axis connecting the footpoints, and rotation about the 
local normal.

\subsection{Ray-marching}\label{Sec:RayMarch}

The standard approach for computing the emission map of a three-dimensional volume
is to use ray tracing, however this becomes significantly more
computationally demanding as the number of discrete volume elements (voxels, 
assumed quasi-homogeneous three-dimensional cubic volumes) increases due to the number of ray-voxel intersection tests that need to be performed to determine whether a certain voxel interacts with a ray.
This method can be optimised by using voxel culling methods, such as octrees, which describe layers of nested voxels, allowing many of the tests against the most refined voxels to be avoided.
These optimisations are quite complex to implement, and given the simplicity of this problem are unnecessary.

Ray-marching instead assumes that the emission and absorption properties of the volume can be sampled continuously.
Instead of computing the radiative transfer integral along a ray between voxel intersections whilst assuming homogeneous plasma parameters, when ray-marching the integral is computed without a grid, by moving a short step along the ray and sampling the local emission and absorption coefficients, whilst assuming homogeneity over this short step. 
If these steps taken along the ray are sufficiently short then this method will tend towards the true value of the integral.

Under the basic description of ray-marching given above, it is assumed that the emission and absorption coefficients of the plasma are defined continuously and can be sampled at any point.
Due to the high cost of computing these coefficients for gyrosynchrotron radiation (Sec. \ref{Sec:GS}) it is not feasible to recompute these at every step along the ray.
The plasma emission and absorption coefficients for the two gyrosynchrotron modes and the thermal radiation are therefore computed on a discrete three-dimensional Cartesian grid and stored in a volume texture (three-dimensional cuboidal array) before the ray-marching procedure.
The coefficients are also multisampled i.e. computed for multiple points
within each voxel and then averaged to attempt to better capture the average plasma conditions  than simply the conditions at the central point.

There are multiple methods for determining when the ray-marching step size needs to be decreased or can be increased.
When the absorption and emission coefficients are continuously defined then metrics based on the local gradient can be used.
In \emph{Thyr} we choose to perform the refinement manually.
The bounds of the volume texture containing the plasma coefficients define an axis-aligned bounding box (AABB) for the object stored in the texture.
Then, using the simple but highly optimised, ``slab'' algorithm
\citep{Kay1986} to compute the intersection of rays with this AABB we have the start and end points between which to ray-march.
This ``slab'' algorithm has been further optimised by relying on strict IEEE754
floating point behaviour.
The manually specified regions of refinement are converted into AABBs contained within the initial AABB.
The plasma coefficients are computed on a finer grid (with user specified refinement factor) within these sub-domains.
If a ray intersects with one of these higher resolution regions (determined by the intersection of the ray with the sub-domain's AABB) then the step size is decreased and the plasma coefficients are sampled from the finer grid when computing the radiative transfer integral.



\textit{Thyr} uses three volume textures to store the coefficients for 
the dipole volume. 
The primary texture represents the entirety of the 
volume at the (lower) coronal
resolution, whilst the other two are heavily refined on the footpoints of the dipole
encompassing the photosphere, chromosphere, and the transition region as 
the atmospheric parameters and magnetic field change much more rapidly in this region.
Whilst a dipole model is chosen for 
our example, any volumetric function or data (such as the results of a magnetogram extrapolation) can be used to
fill the parameters of the texture, and the ray-marching approach will remain
unchanged.
The model is currently based around the concept of a single AABB containing the
entire scene at low-resolution and a number of refined regions contained
strictly within this AABB. 
The information from the low-resolution AABB is then ignored in the locations 
where a refined region is present.
These AABBs serve as the boundaries of a two-level three-dimensional Cartesian grid hierarchy, with a uniform grid within each level.
If multiple low resolution AABBs are desired this change would be relatively
trivial, but using the convention of a single low resolution AABB allows any
geometry to be traced by \textit{Thyr} with no changes to the ray-marching
code.
It would equally be simple to extend the code to support a full multi-level grid hierarchy with further refined regions within the refined regions, if extreme resolution were required in some locations.

Fig.~\ref{Fig:RT} shows (in two dimensions) the difference
between the ray-tracing and ray-marching.
The cost of finding the emission and absorption parameters for a point on the ray is 
constant if these parameters are described by a simple expression or discretised onto a known Cartesian grid, as these parameters can be computed or looked-up wherever the blue point happens to be.
When applying a traditional, simple, ray-tracing technique without additional voxel culling optimisations, the ray-voxel intersection test must be performed against every voxel in the domain to determine the locations of the green points.
This computation scales linearly with the number of voxels.
Therefore, using the ray-marching approach, it is computationally tractable to 
sample the volume with a step size significantly smaller than the voxel side length.
The integral then amounts to simply performing several elementary mathematical 
operations for each step to compute the closed form radiative transfer integral from a homogeneous slab, and looking up one value in a large array.
Due to the cost of calculating the many ray-voxel intersections, for the
cases used in \textit{Thyr} ray-marching is typically 1.5 orders of magnitude
faster than ray-tracing, even though the integral of radiative transfer is computed at
significantly more points including a higher resolution region (orange box in Fig.~\ref{Fig:RT}).

An assumption that is implicit to the ray-marching approach is that it is 
acceptable to not sample voxels for which the space between the intersections is 
very small (e.g. just cutting through close to a corner), as their effect is insignificant.
In practice with a sufficiently small step-size (\emph{Thyr} uses 0.1$\times$ the local grid side length for the plasma coefficients) 
and reasonably smoothly varying emission and absorption coefficients this is not a 
problem as the effect of the contribution from this region is not significant.
The multisampling of the coefficients helps improve the smoothness of the 
coefficients between voxels and guarantee that the parameters chosen accurately 
reflect the plasma contained within (assumed homogeneous).

\section{Verification against Klein \& Trottet (1984)}

   \begin{figure}
   \centering
   \includegraphics[width=1.05\hsize]{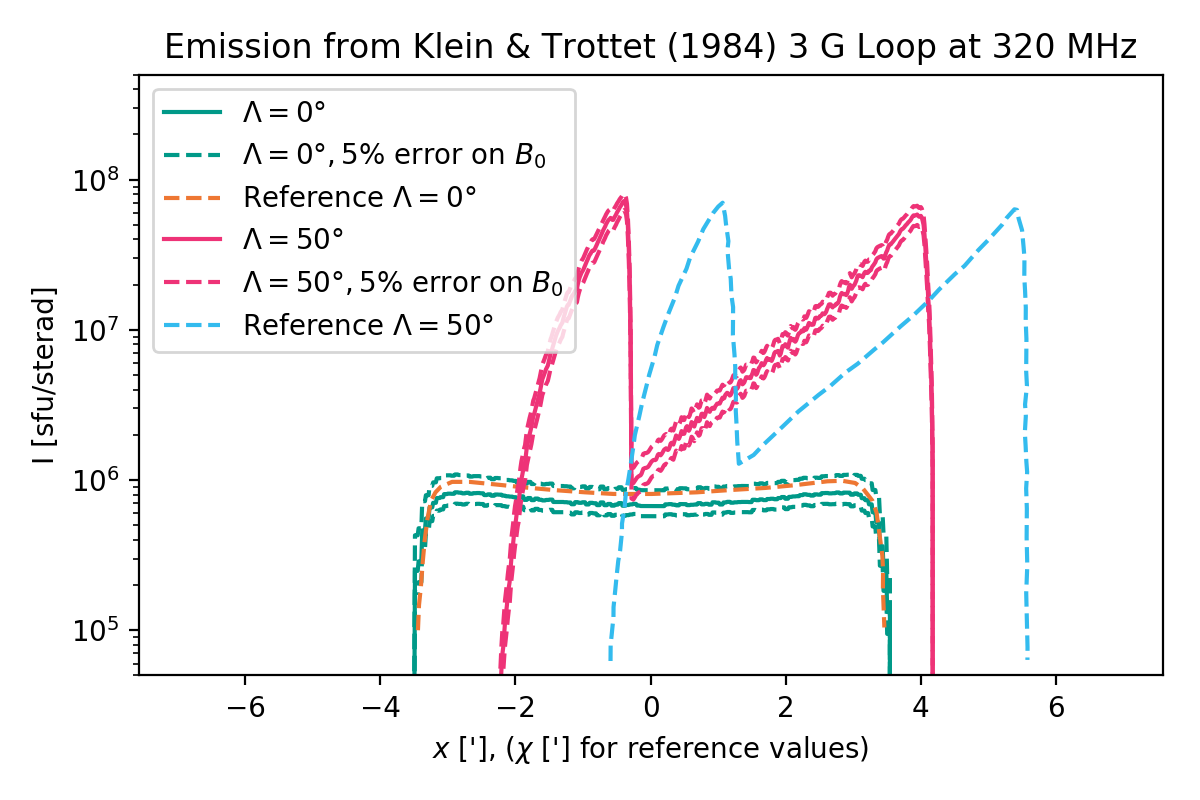}
      \caption{Valdation of \emph{Thyr} against the model of \citet{Klein1984} (reference values)}
         \label{Fig:Klein}
   \end{figure}

To verify \textit{Thyr}'s correctness we use a problem presented in \citet{Klein1984}.
In this test case the gyrosynchrotron emission from a two-dimensional 
loop is simulated. 
The loop is observed from two different viewing angles $\Lambda$. 
The specific intensity (sfu\footnote{solar flux units 
(1 sfu = $10^4$ Jy = $10^{-22}$ W m$^{-2}$ Hz$^{-1}$ = $10^{-19}$ erg s$^{-1}$ cm$^{-2}$ Hz$^{-1}$)}/sterad) 
is then computed for a telescope beam scanning across the loop with angle $\chi$. 
In \emph{Thyr} the rays along the observer's line of sight are assumed parallel, and do not 
diverge from a point like the telescope in \citet{Klein1984}. 
The coordinate $x$ is then the displacement in \emph{Thyr's} imaging plane in units of
arcminutes on the surface of the Sun viewed from Earth.

Using Klein and Trottet's parameters for loop with 3~G coronal magnetic field strength
at $\Lambda=0^\circ$ and $\Lambda=50^\circ$ we have
reproduced the emission at 320 MHz in Fig.~\ref{Fig:Klein}.
The results are very similar in shape and intensity.
The slight differences in intensity can be accounted for by modifying the looptop magnetic 
field strength by less than 5\%.
\citet{Klein1984} define the looptop magnetic field strength at the outer boundary of 
the looptop, whereas in \emph{Thyr} it is defined in the centre.
This value had to be manually tuned to obtain a magnetic field strength similar to 
that used by \citet{Klein1984} in the slice taken from the three-dimensional simulation performed in \emph{Thyr}.

The spatial offset of \emph{Thyr's} results in the $\Lambda=50^\circ$ case is due to \emph{Thyr} 
performing the rotation but also translating the dipole so as to keep it centered in the field of view.

\section{Example 3D Simulations}

To demonstrate \textit{Thyr's} capabilities we performed a set of flare simulations 
using a simple dipole model for the structure of the emitting volume and the magnetic field, 
a power-law non-thermal electron distribution, and the other atmospheric parameters 
specified by a semi-empirical model. 
From these simulations we obtain brightness maps, spectra, and polarisation maps.
Complete spectral data is available for every pixel and region of these maps, 
thus allowing us to perform a localised spectroscopic analysis (Section~\ref{Sec:ImagingSpectro}). 
We perform these simulations both with and without resolving the chromosphere in detail to show its role on emission at higher frequencies.
These simulations can serve as examples for how to set up and run the code, in addition
to basic post-processing of results.

\subsection{Semi-empirical atmosphere and parameters}\label{Sec:Atmosphere}

\begin{figure}
\centering
\includegraphics[width=\hsize]{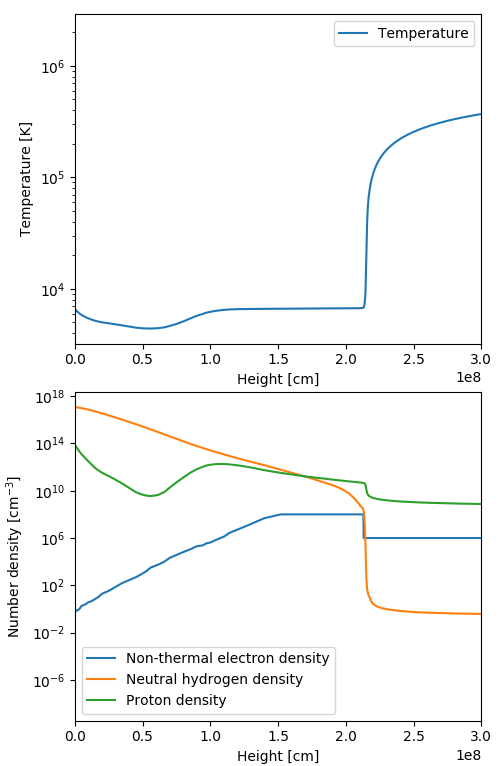}
   \caption{Atmospheric parameters in our C7 \citep{Avrett2008} based model. \textit{Top:} Temperature. \textit{Bottom:} Number density of protons, neutral hydrogen and non-thermal electrons (see text).}
      \label{Fig:C7Atm}
\end{figure}

In our example simulations we use the C7 quiet Sun semi-empirical atmosphere
of \citet{Avrett2008}.
This atmosphere is extrapolated to coronal parameters, by extrapolating linearly in log-log space the parameters from the top of the C7 model (at 47 Mm), until 150 Mm where the atmosphere is then set constant. Our atmosphere is plotted in Fig.~\ref{Fig:C7Atm}.

A non-thermal electron distribution $g(E, \phi)$ is embedded in the semi-empirical atmosphere, with a
density defined as $n_e=10^6$ cm$^{-3}$ in the corona, and $n_e=10^8$ cm$^{-3}$ at
the top of the chromosphere. 
This increase of $n_e$ is caused by the decrease of the emitting volume 
due to the convergence of the dipole magnetic field.
As can be seen Fig.~\ref{Fig:C7Atm}, we have made the approximation to make this increase 
a step rather than scaling directly with the area of the dipole.
This decision was made to ensure a simple example using a dipole magnetic field that produces chromospheric radio emission and to provide an electron number flux of $~10^{36}$ s$^{-1}$ entering the
chromosphere, which is consistent with observations \citep[e.g.][]{Simoes2013}.
These electrons follow a power-law distribution in energy, with a minimum cut-off
energy of $E_\mathrm{min}=10$ keV, a maximum cut-off of $E_\mathrm{max}=5$ MeV, and a spectral index of $\delta=3$. 
We assume that these electrons have an isotropic pitch-angle distribution,
but the code supports arbitrary pitch-angle distributions.

As we descend into the chromosphere the non-thermal electron distribution is
attenuated by balancing collisional losses against the column density
traversed by the electrons.
We use the approximate relationship that the critical stopping column density is $N_\textrm{stop}\approx
10^{17}E_0^2$ where $E_0$ is the initial electron energy in keV and
$N_\textrm{stop}$ is in cm$^{-2}$ \citep{TandbergHanssen1988}.
The lower energy bound of the power law is then proportionately increased as 
electrons below the stopping energy are cut off.

Finally, for these models we use a dipole with height of approximately
$3.86\times10^9$ cm ($53.3"$) and a centre-to-centre footpoint separation of
$2.57\times10^9$ cm ($35.5"$).
The radius at the looptop ($\rho_0$ in the \citet{Kuznetsov2011} model) is $7.72\times10^8$ cm.
The looptop field is set to 100 G, yielding a strength of the order of 2000 G
at the photosphere.
The dipole here has a latitude of 30$^\circ$, longitude of 70$^\circ$, and rotation 
about the local normal of $-20^\circ$. 

The simulations presented here are performed both with and without refined
resolution in the chromosphere.
In the low resolution simulation the voxels have a side length of
1800 km.
In the high resolution simulation the voxels of the upper corona have a 
450~km side length and the refined voxels in the lower atmosphere
have a 75~km side length (refined six times).
\emph{Thyr's} ray-marching algorithm automatically transitions between the low- 
and high-resolution regions with no ill effects due to the boundary.
An additional high resolution simulation was also computed with voxels with 100~km side lengths, and no significant spectral difference was found between the two, so it was concluded that the 75~km resolution was sufficient to capture the details of the problem.

\subsection{Emission Maps}\label{Sec:EmissionMaps}

   \begin{figure*}
   \centering
   \begin{tikzpicture}
     \draw (0, 0) node [inner sep=0] {\includegraphics[width=0.95\columnwidth]{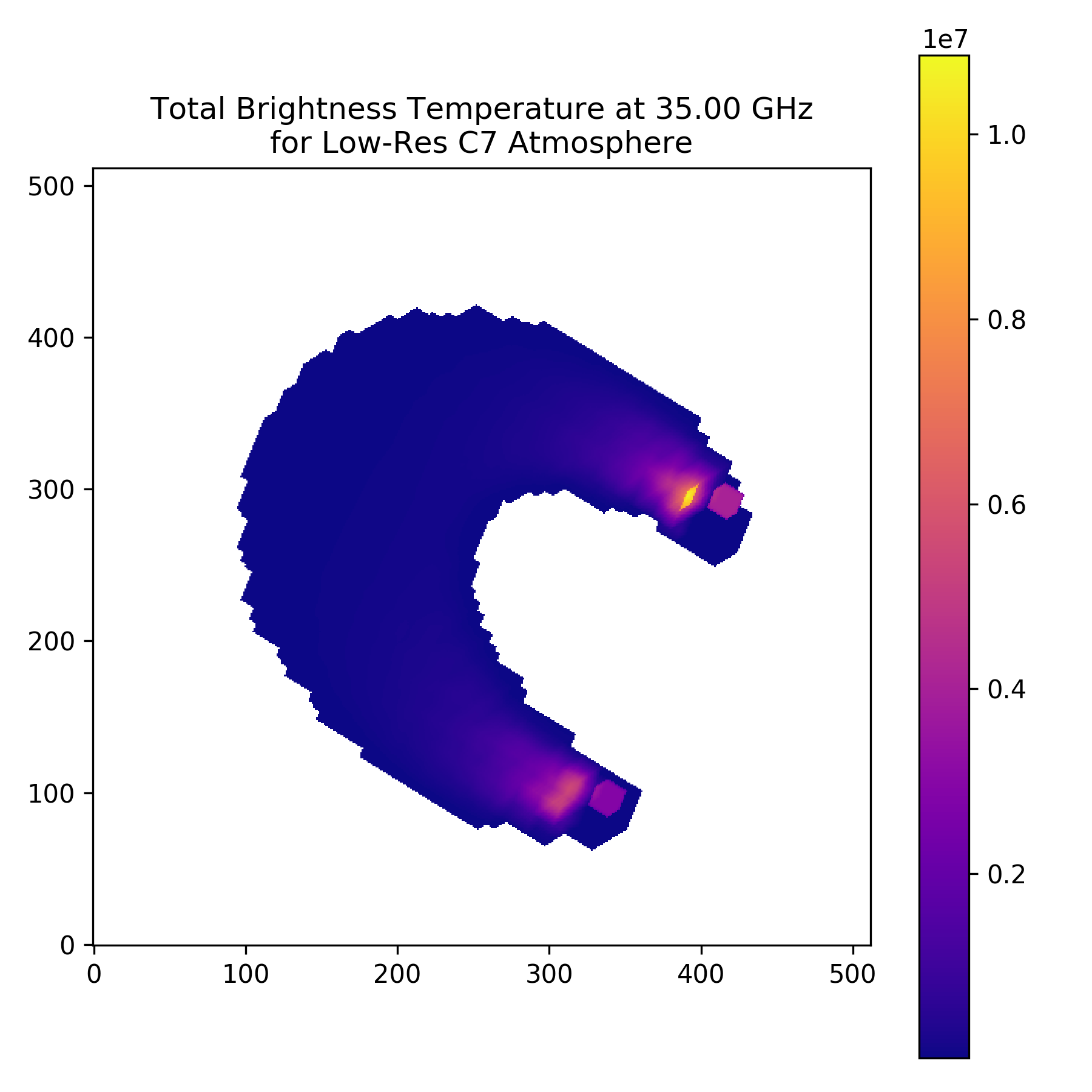}};
     \draw (-0.4\columnwidth, 0.4\columnwidth) node {a)};
   \end{tikzpicture}
   \begin{tikzpicture}
     \draw (0, 0) node [inner sep=0] {\includegraphics[width=0.95\columnwidth]{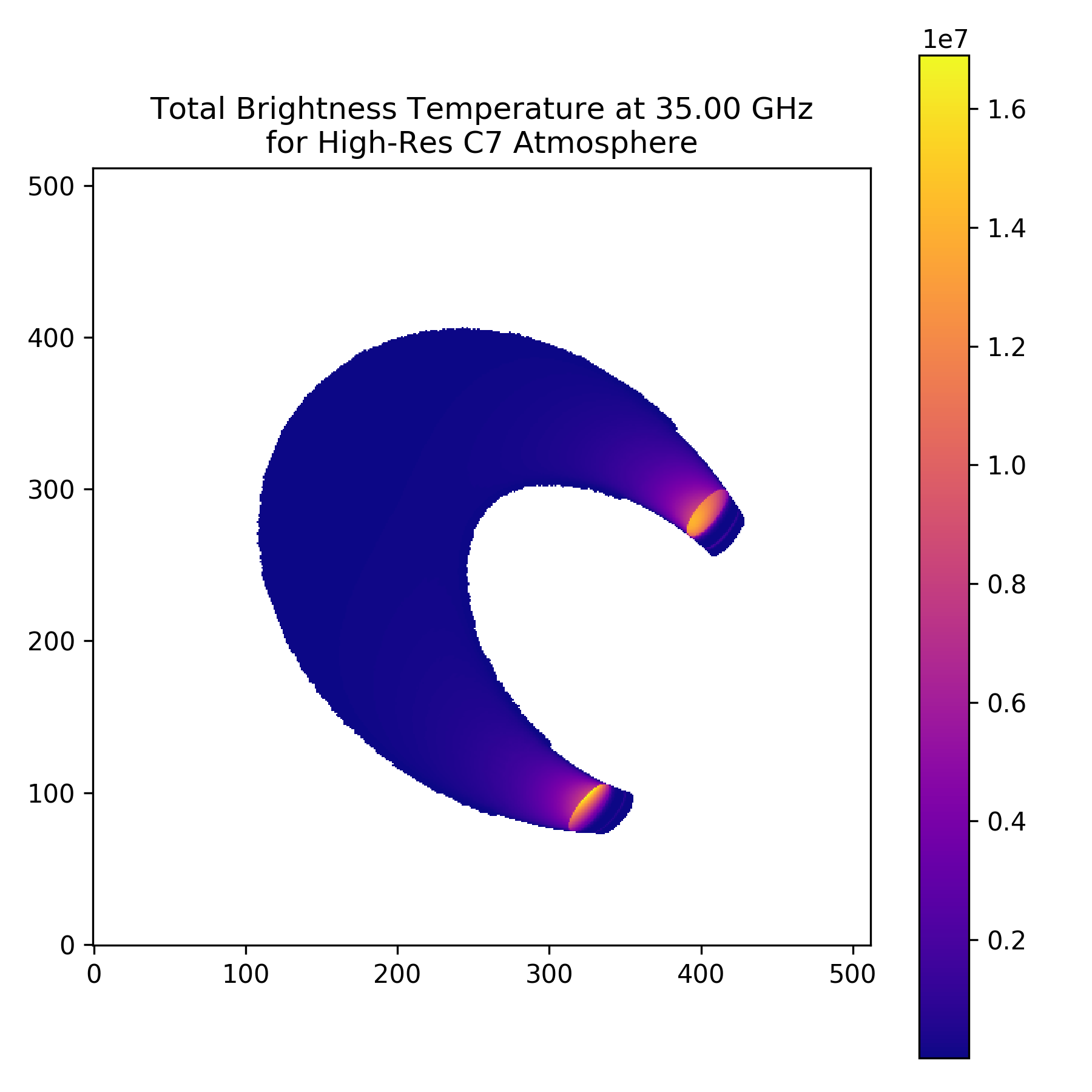}};
     \draw (-0.4\columnwidth, 0.4\columnwidth) node {b)};
   \end{tikzpicture}
   \caption{Total Brightness Temperature at 35 GHz for a loop simulated using only low resolution voxels (a))
   and a loop with increased resolution in the lower corona and chromosphere (b)). 
   Note the increased peak brightness temperature between b) and a). Nearly all the emission at this frequency ($T_b \gtrsim 5$ MK) originates from the refined region, which can be seen in b) by the slight decrease in radius and smoother curvature.}
   \label{Fig:C7Tb}
   \end{figure*}

\begin{figure}
\centering
\includegraphics[width=\hsize]{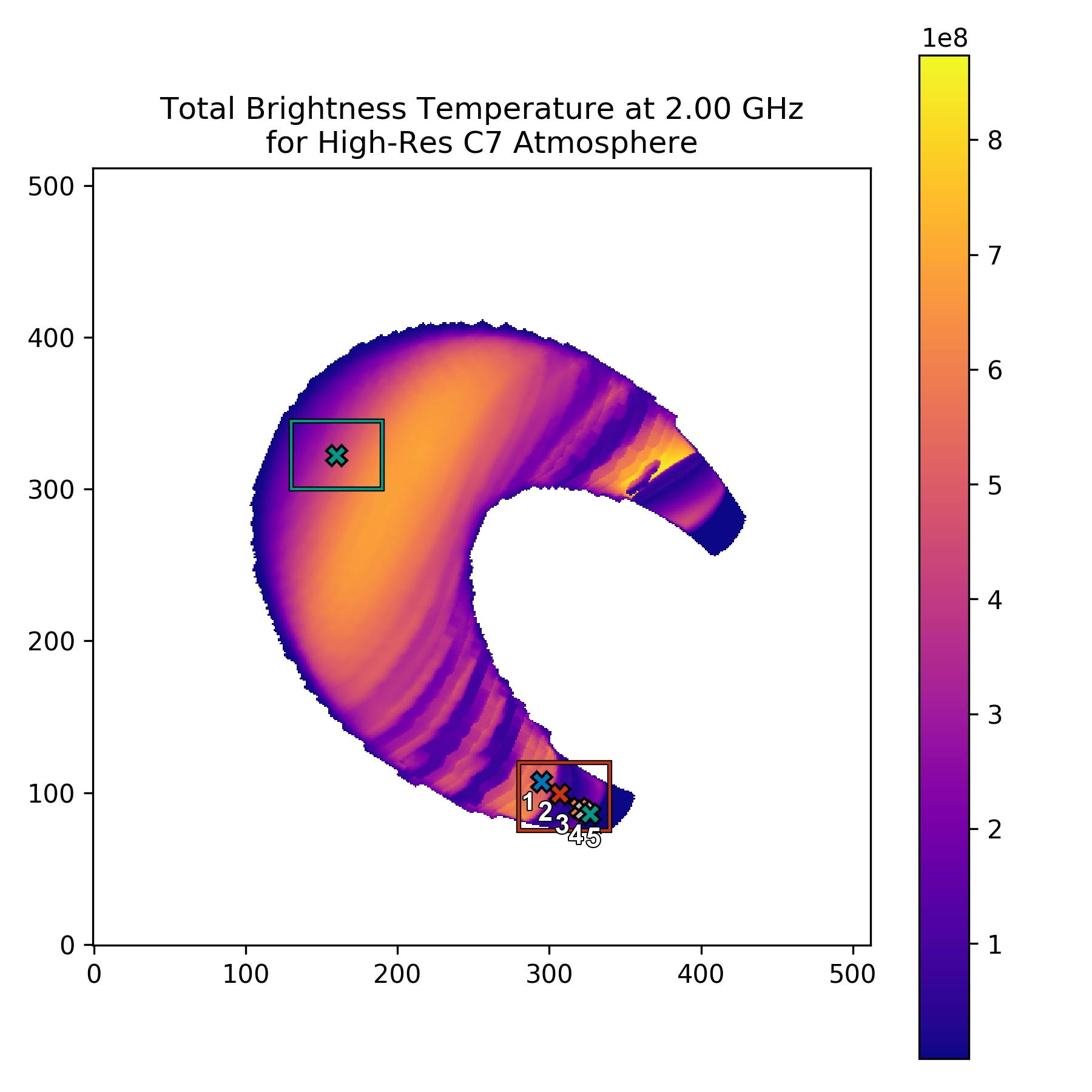}
  \caption{Total Brightness Temperature at 2 GHz for a high-resolution loop.
  The plasma is optically thick at this frequency thus showing the complex 
  structure of the gyrosynchrotron emission. There is a clear boundary between 
  the optically thick emission in the lower section of the loop and the optically 
  thin emission in the looptop.
  This boundary moves downwards towards the footpoints as the observed 
  frequency increases.
  The regions and points marked on the emission map are used 
  in Section~\ref{Sec:ImagingSpectro}}
  \label{Fig:OptThick}
\end{figure}

Fig. \ref{Fig:C7Tb} shows the total brightness maps at 35~GHz for the
same loop simulated at different resolutions.
The emission from the high-resolution loop at 2 GHz is shown in Fig.~\ref{Fig:OptThick}.
This shows the complex nature of gyrosynchrotron emission at frequencies at 
which the plasma is optically thick.
The high frequency emission maps (Fig.~\ref{Fig:C7Tb}) have the most intense 
emission from the footpoints, associated with the strongest magnetic field 
values. 
In the low frequency emission maps (Fig.~\ref{Fig:OptThick}) the stripes are 
caused by the harmonic structure of the gyrosynchrotron emission at a single 
frequency for a spatially varying $B$ field.
The effects of insufficient resolution within the simulation can be seen by
comparing the simulations in Fig.~\ref{Fig:C7Tb}.
The low-resolution simulation of Fig~\ref{Fig:C7Tb}a 
produces an lower brightness temperature than the high-resolution model shown in Fig.~\ref{Fig:C7Tb} due to the small region this high brightness originates from.

\subsection{Spectra}

\begin{figure*}
\centering
   \begin{tikzpicture}
     \draw (0, 0) node [inner sep=0] {\includegraphics[width=0.95\columnwidth]{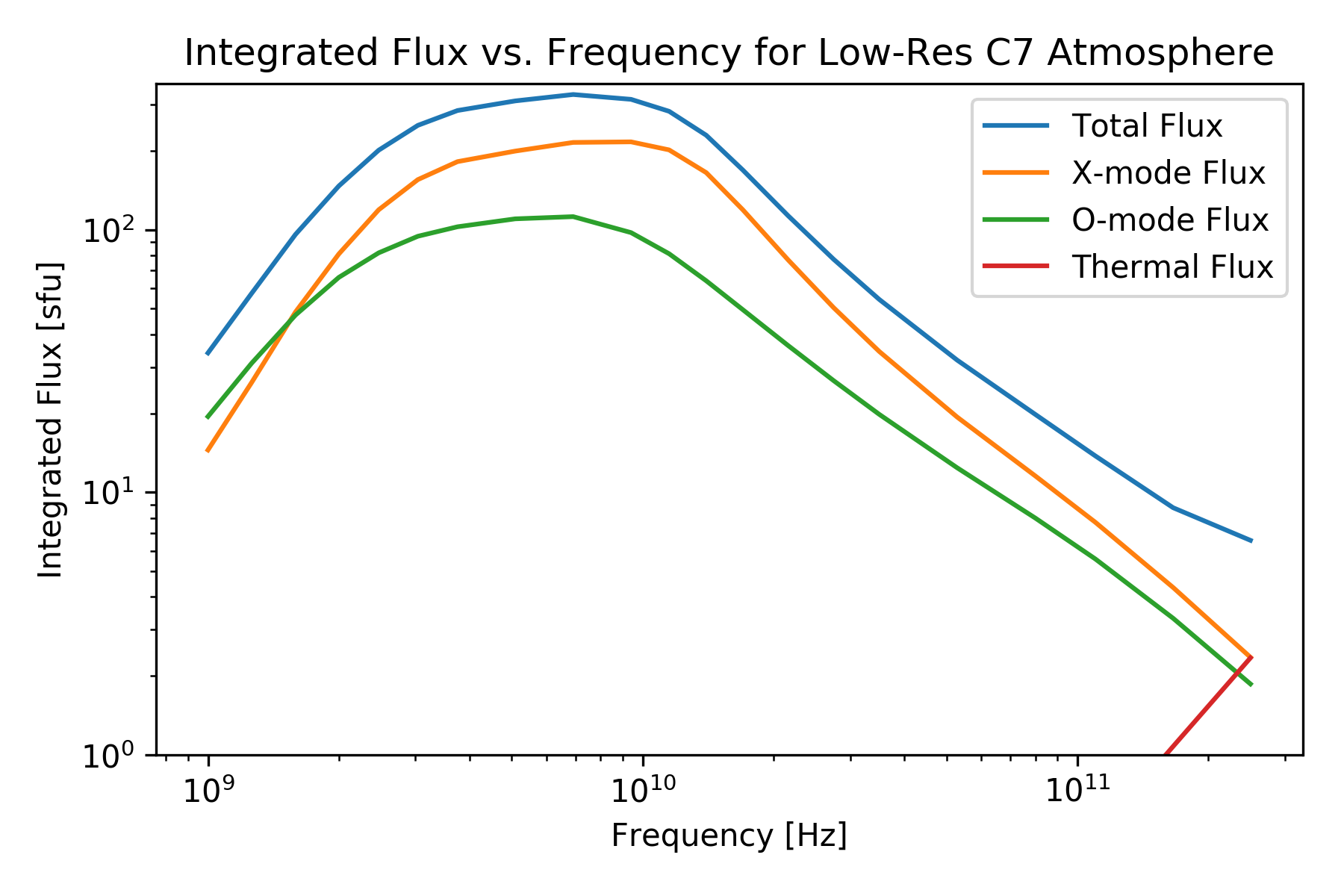}};
     \draw (-0.4\columnwidth, 0.267\columnwidth) node {a)};
   \end{tikzpicture}
   \begin{tikzpicture}
     \draw (0, 0) node [inner sep=0] {\includegraphics[width=0.95\columnwidth]{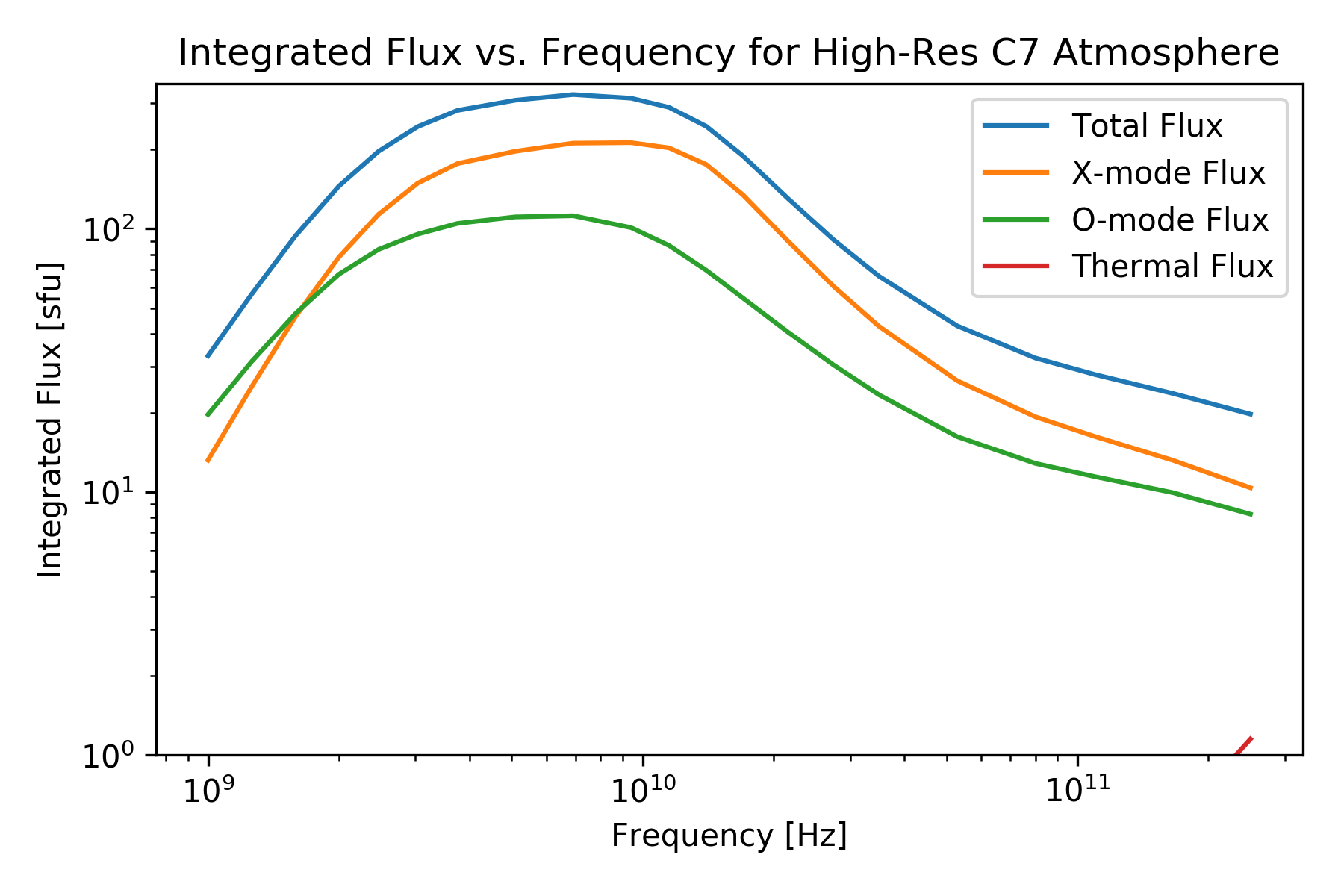}};
     \draw (-0.4\columnwidth, 0.267\columnwidth) node {b)};
   \end{tikzpicture}
  \caption{Spectrum of integrated flux from the flaring loop, using the low-- (a)) and high--resolution (b)) model.}
      \label{Fig:C7Flux}
\end{figure*}

Integrating over the brightness temperature maps, the flux density spectra 
$F_\nu$ of these two simulations can be computed. 
In practical terms, this calculation is simply the sum of the flux density 
of each pixel (column $i$ and row $j$), obtained from their specific 
intensity $I_{ij}$ over the pixel solid angle $\Omega_\mathrm{pixel}$:
\begin{equation}
F_\nu = \sum_{i=1}^{N_x} \sum_{j=1}^{N_y} I_{ij} \ \Omega_\mathrm{pixel}.
\end{equation}
The specific intensity $I_{ij}$ of the pixels are directly found from their 
brightness temperature values $T_{b,ij}$ via the Rayleigh-Jeans law:
\begin{equation}
I_{ij} = \frac{2k_b \nu^2}{c^2} T_{b,ij}
\end{equation}

The resulting spectra are shown in Fig.~\ref{Fig:C7Flux}, 
in sfu. 
Both of these plots show the characteristic ``inverted-v'' shape of
gyrosynchrotron emission, and the fact that the ordinary mode dominates at
low frequencies and the extraordinary mode dominates at higher frequencies \citep{Ramaty}.
The thermal emission is plotted, but is insignificant in the cases we are
investigating here.
The difference in the magnitude of the thermal emission is due to a region that 
is small and bright in the high-resolution case, being spread across a much larger 
voxel in the low-resolution case.
Although the spectrum of gyrosynchrotron radiation from a uniform source would 
present harmonics, they are not visible in the spatially-integrated spectra due to 
the overlap of harmonic structures from a spatially-varying $B$ field and atmospheric parameters,
as has been previously discussed by \citet{Klein1984,Simoes2006}.
The spectra for the low- and high-resolution cases behave similarly at low
frequencies, but differences emerge at higher frequencies.
In the high-resolution simulation there is a hardening of the spectrum of the ordinary 
and extraordinary modes of the gyrosynchrotron radiation at high frequencies.
As these non-thermal modes dominate over the thermal emission the hardening in the
total spectrum must be non-thermal in origin, as we will discuss in Section~\ref{Sec:Discussion}.

\subsection{Polarisation Degree}

\begin{figure}
\centering
\includegraphics[width=\hsize]{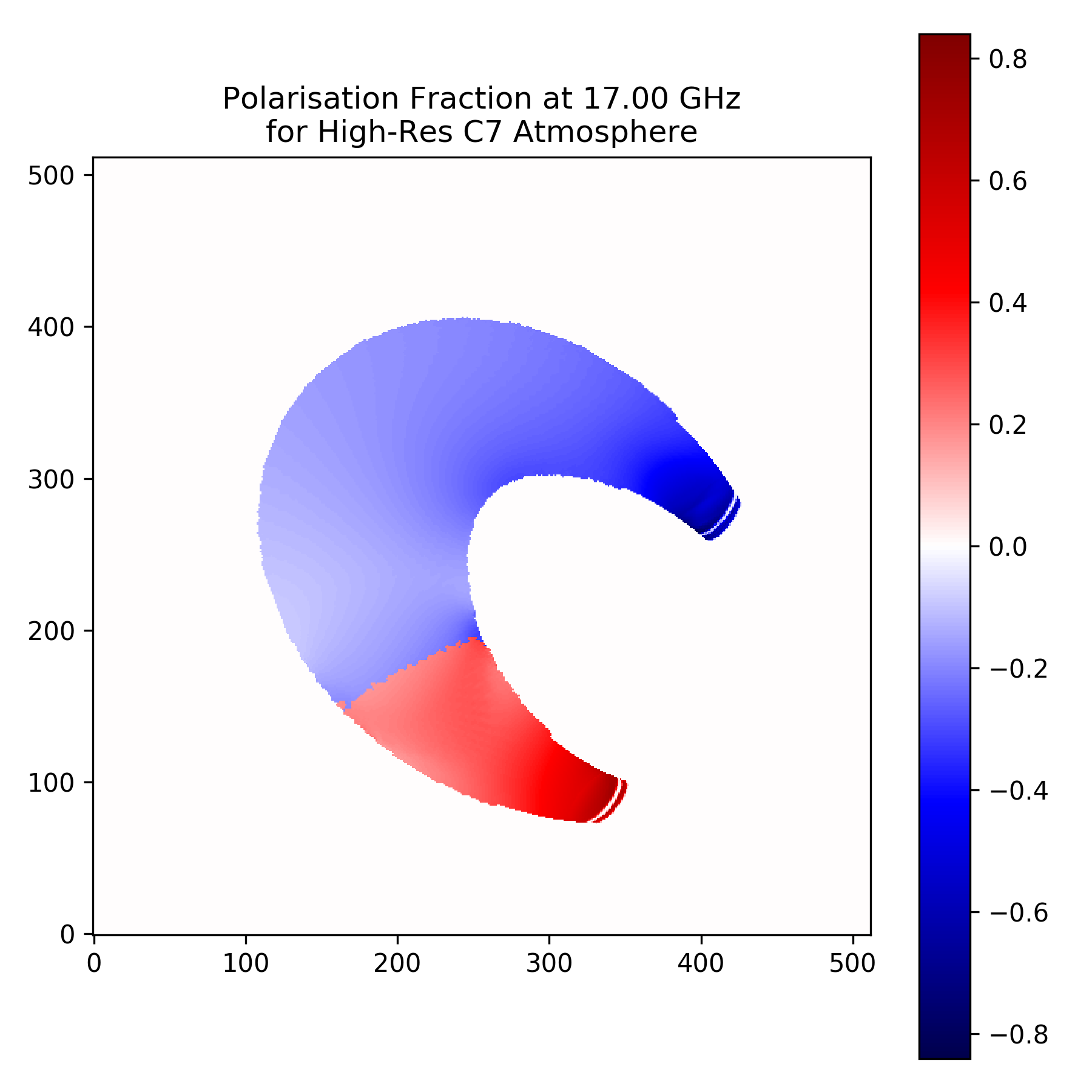}
  \caption{Polarisation degree for the high resolution simulation at 17 GHz.}
      \label{Fig:C7Pol}
\end{figure}

Fig.~\ref{Fig:C7Pol} shows a map of the polarisation degree for the high
resolution model at 17 GHz.
The polarisation degree is given by the Stokes parameters $V/I$.
The expected opposite polarisation in each footpoint is present.
The asymmetry in peak polarisation degree between the two footpoints is
purely an effect of viewing angle -- the loop is fully symmetric.
\citet{Klein1984} and \citet{Simoes2006} have previously shown the importance 
of the viewing angle on observed polarisation, and our results are in accordance.
With a three-dimensional system there is a variety of situations that can 
produce significant differences between the intensity of footpoints, 
and the location of the change in polarisation direction.
Polarisation data is an important component of radio observation, one 
that could be used to shed light on the magnetic field geometry 
\citep{Simoes2006}, and an essential tool for diagnosing 
the electron energy distribution \citep{Kuznetsov2011}.

The structure of the polarised radiation shown in Fig.~\ref{Fig:C7Pol} is 
quite simple with a single transition between negative and positive polarisation.
At lower frequencies where a large portion of the volume is optically 
thick and the complex ``gyrostripe'' patterns (see Fig.~\ref{Fig:OptThick}) 
are visible the polarisation patterns are also far more 
complex and follow these stripes.

In the simulations presented here a simple dipole magnetic field model is used.
As the polarisation degree is strongly related to the direction of the 
magnetic field a more complex magnetic model can yield very different 
polarisation patterns \citep[e.g.][]{Gordovskyy2017}. 
\textit{Thyr} is capable of using an arbitrary magnetic field geometry 
and thus can be used to investigate cases with complex polarisation patterns, 
including fine structure if the resolution is increased in the region of interest.


\subsection{Imaging Spectroscopy}\label{Sec:ImagingSpectro}

\begin{figure*}
\centering
   \begin{tikzpicture}
     \draw (0, 0) node [inner sep=0] {\includegraphics[width=0.95\columnwidth]{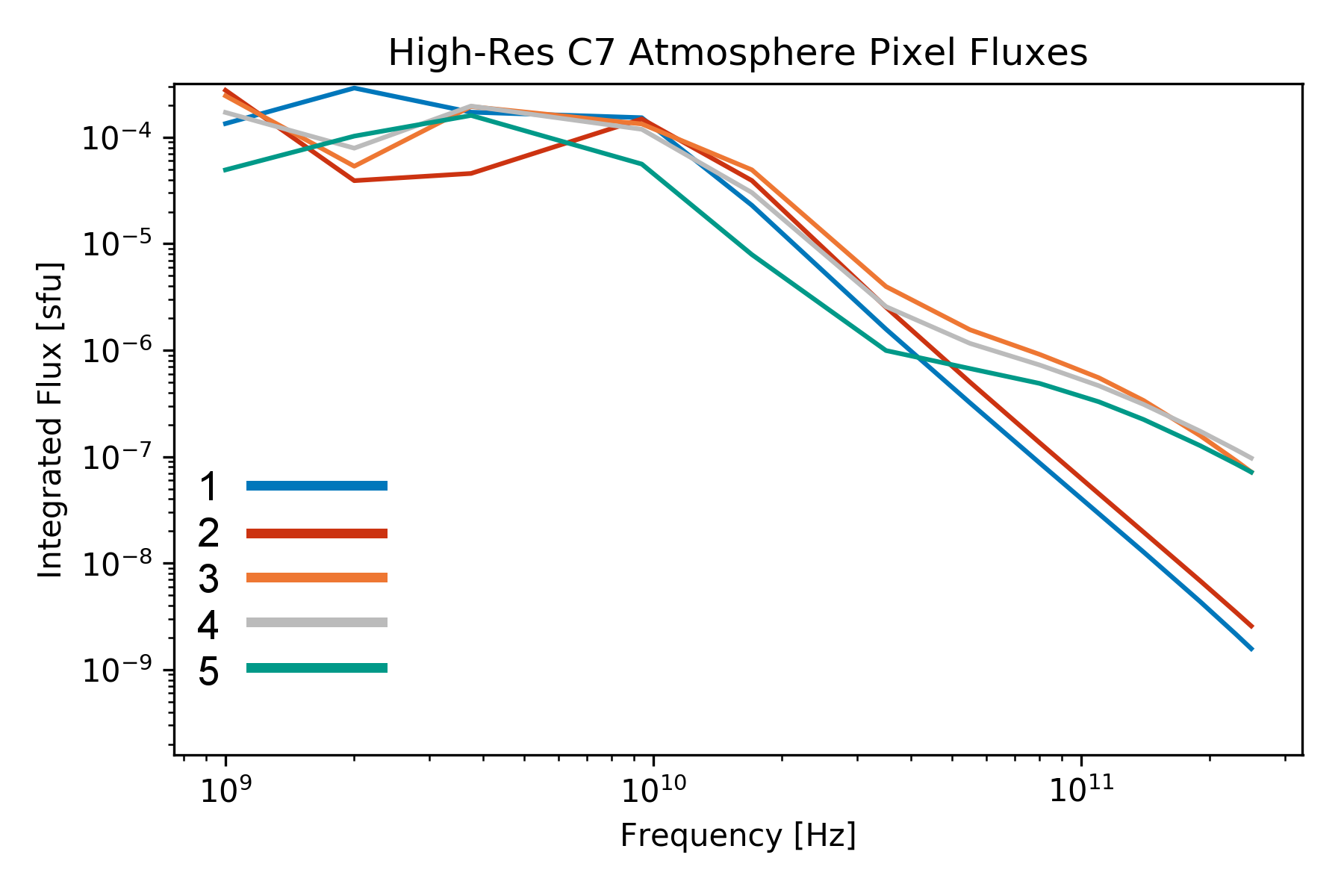}};
     \draw (-0.4\columnwidth, 0.267\columnwidth) node {a)};
   \end{tikzpicture}
   \begin{tikzpicture}
     \draw (0, 0) node [inner sep=0] {\includegraphics[width=0.95\columnwidth]{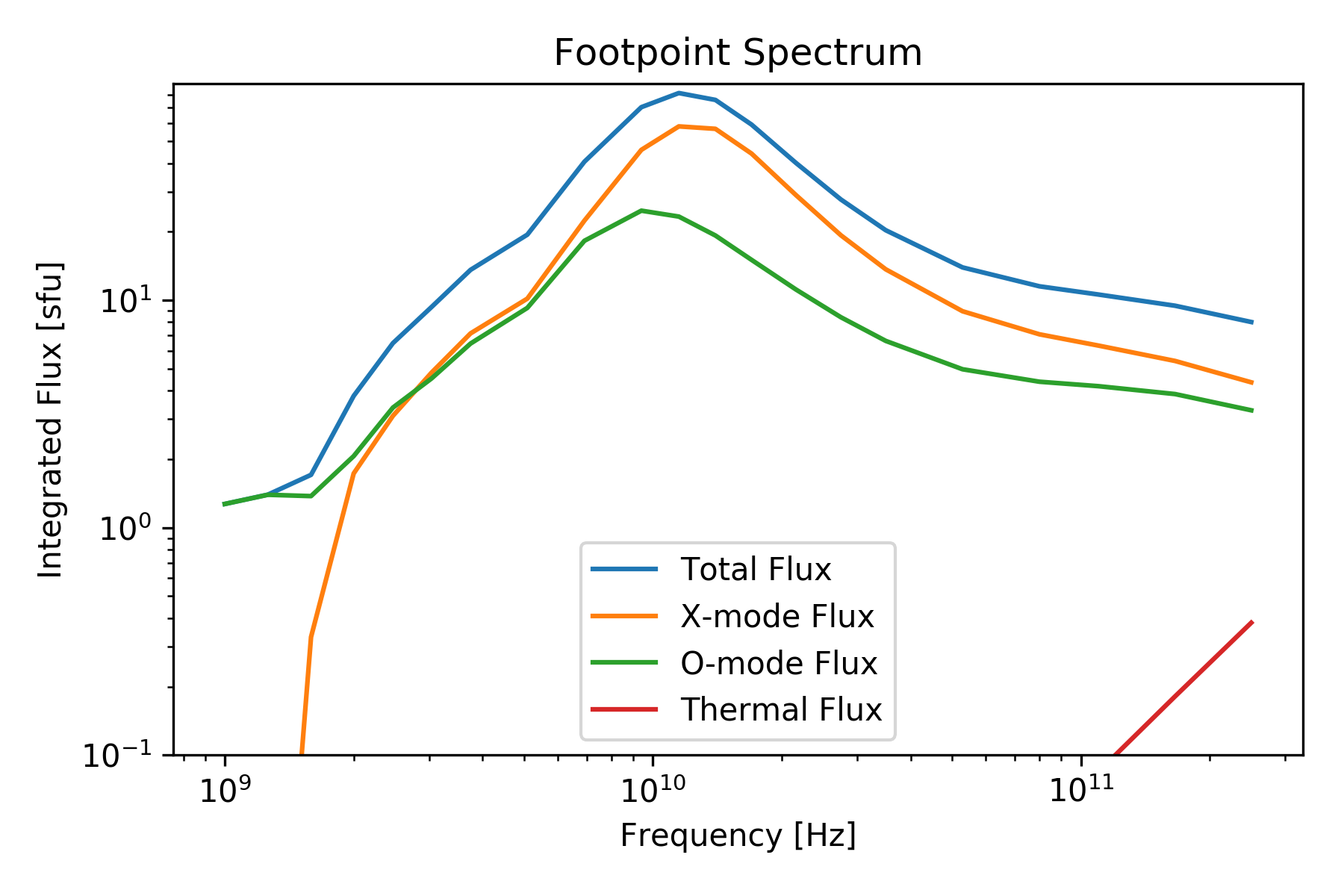}};
     \draw (-0.4\columnwidth, 0.267\columnwidth) node {b)};
   \end{tikzpicture}
   \begin{tikzpicture}
     \draw (0, 0) node [inner sep=0] {\includegraphics[width=0.95\columnwidth]{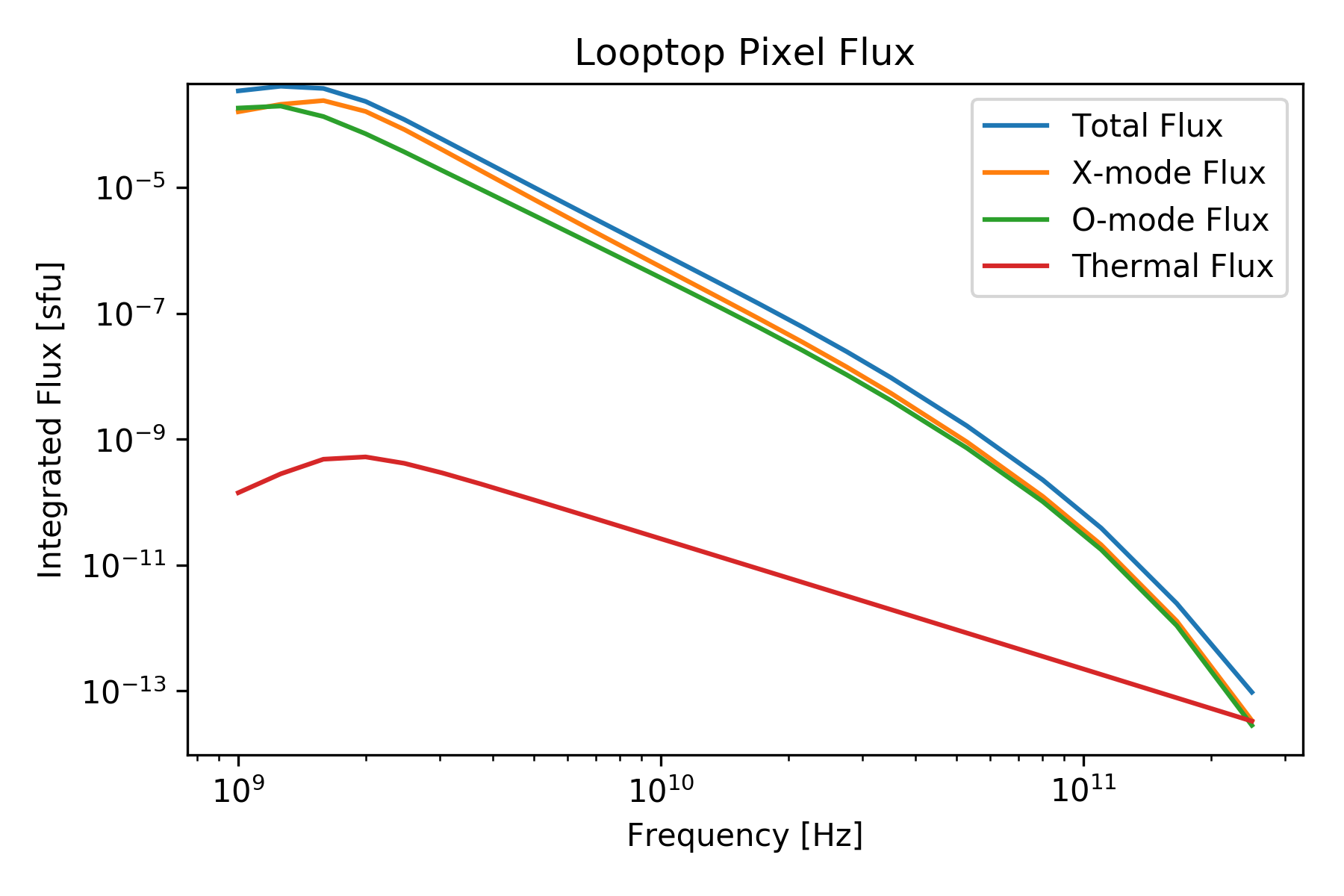}};
     \draw (-0.4\columnwidth, 0.267\columnwidth) node {c)};
   \end{tikzpicture}
   \begin{tikzpicture}
     \draw (0, 0) node [inner sep=0] {\includegraphics[width=0.95\columnwidth]{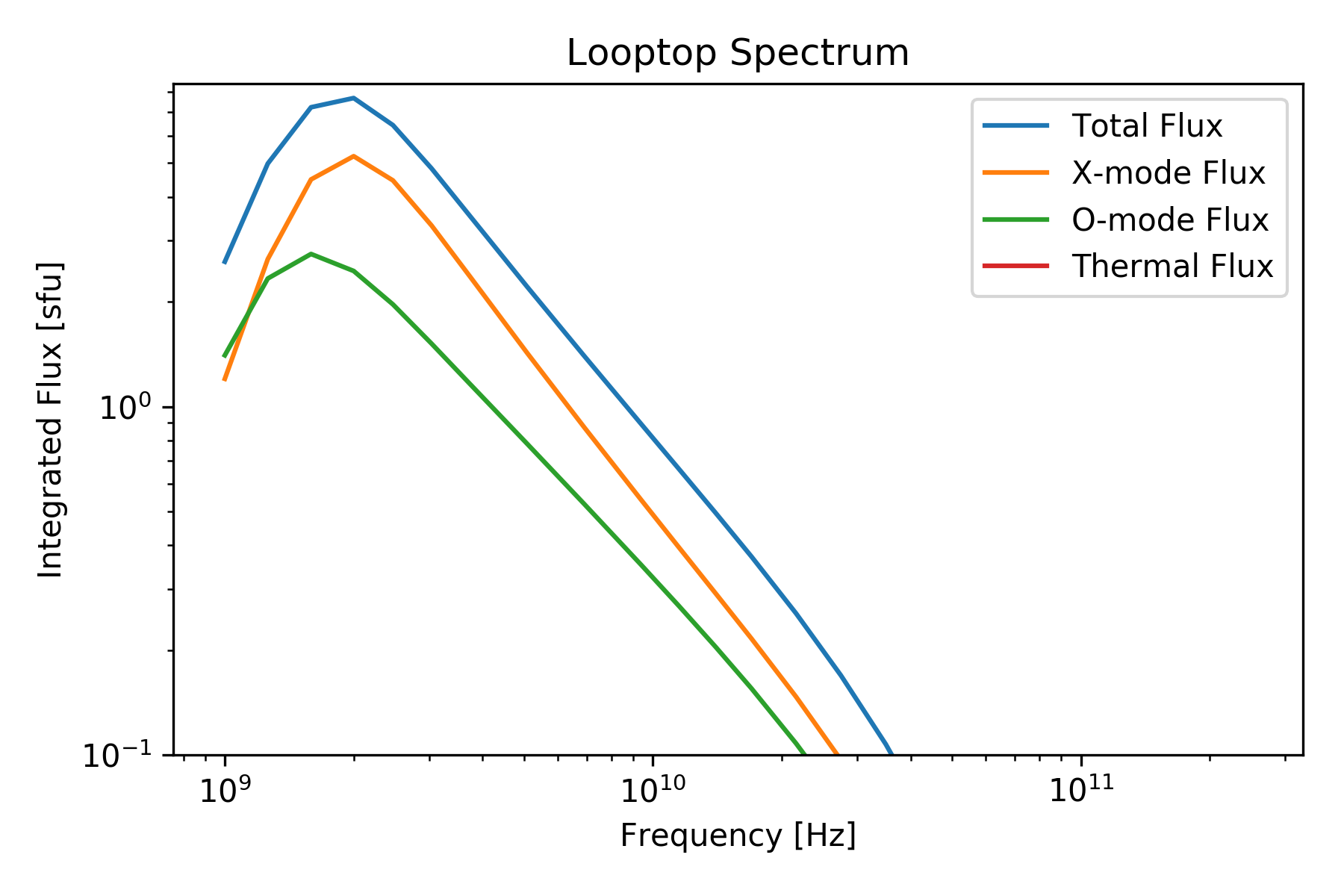}};
     \draw (-0.4\columnwidth, 0.267\columnwidth) node {d)};
   \end{tikzpicture}
  \caption{Spectra of the marked footpoint pixels (a)), footpoint box (b)), looptop single pixel (c)), and looptop box (d))}
      \label{Fig:ImagingSpectro}
\end{figure*}

In Fig.~\ref{Fig:OptThick} we have marked a looptop and a footpooint region
in teal and red, respectively.
The central pixel is marked in teal for the looptop region, and 
multiple pixels are marked and numbered inside the footpoint region.
The simulated spectral flux from the marked pixels and integrated over 
the regions is plotted in Fig.~\ref{Fig:ImagingSpectro}, where 
the colours in Fig.~\ref{Fig:ImagingSpectro}a indicate the total 
emission from the pixel of the same colour (not separated by 
emission mode). 
In the footpoint region no extraordinary-mode emission is seen at less than 2 GHz 
as the extraordinary-mode is unable to propagate due to the ratio 
of the plasma frequency to the gyro-frequency being too low 
(see \eqref{Eq:ExtraProp} and \citet{Ramaty}). 
It can be clearly seen that the peak emission occurs at a significantly 
lower frequency in the looptop than in the footpoint. 
Additionally, the peak flux emitted from the footpoint region is much larger 
than that from the looptop.
The high frequency emission from the looptop drops off as expected 
for a homogeneous gyrosynchrotron source.
This is not so for the footpoint region. 
Fig.~\ref{Fig:ImagingSpectro}b clearly shows that the footpoint is 
the origin of this hardening, which is logical as this effect is 
only seen in simulations with a heavily refined lower atmosphere.
By investigating the spectra of the marked pixels within the footpoint region, 
we can see that the peak emission frequency continues to increase 
significantly with depth in the atmosphere, and that pixels 3, 4, and 5 
lie within the compact region of the atmosphere
from which the high-frequency hardening originates.

The results presented here were simulated with an isotropic 
electron pitch-angle distribution.
As described in Section~\ref{Sec:Atmosphere} the lower energy electrons are 
prevented from reaching the lower atmosphere as they are removed from the 
distribution above a critical column density.
If a loss cone style distribution of pitch angles were also applied 
the low--energy end of the distribution could drop off faster 
in the atmosphere, further enhancing this high-frequency hardening in 
the footpooints \citep{Simoes2010,Kuznetsov2011}.
Arbitrary pitch angle distributions are supported by the gyrosynchrotron 
code in \textit{Thyr} and this effect could therefore be easily investigated.

\section{Code Description}

\textit{Thyr} requires a modern C++14 compliant compiler, Torch7 (built on
LuaJIT), and Python 3 with the matplotlib package. 
The core calculation of the gyrosynchrotron components is computed in C++,
whilst all set up, and ray-marching is performed in Lua.
Plotting is done through our \textit{Plyght} tool, also available on Github
\footnote{\url{https://github.com/Goobley/Plyght}},
that allows any language with a socket API or foreign
function interface to easily produce two-dimensional plots using matplotlib \citep{Hunter2007}.
The code also supports exporting all data to the widely supported \textit{comma separated variable} (csv) 
format to allow for post-processing and plotting in any language.

\section{Discussion} \label{Sec:Discussion}

\begin{figure*}
\centering
   \begin{tikzpicture}
     \draw (0, 0) node [inner sep=0] {\includegraphics[width=0.95\columnwidth]{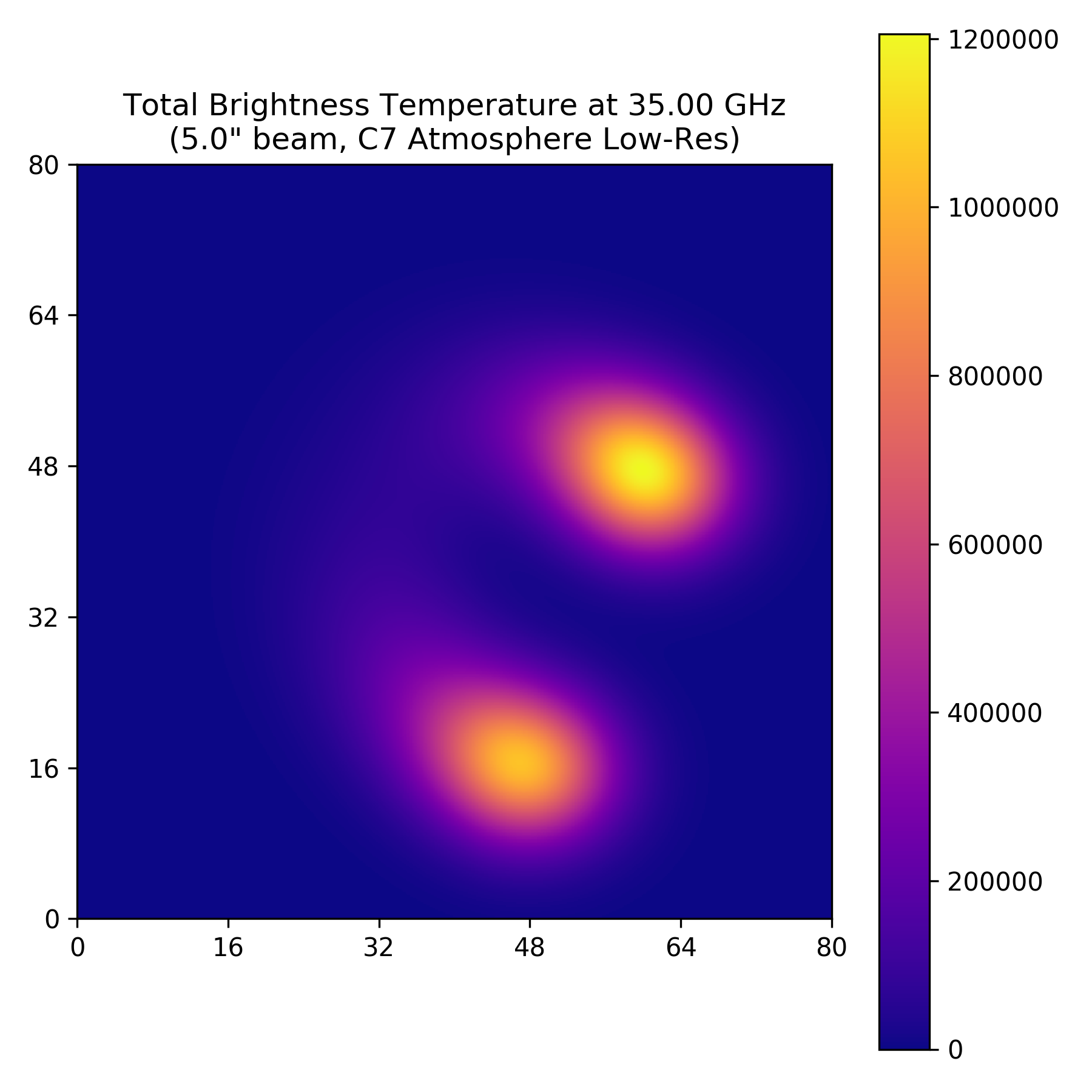}};
     \draw (-0.4\columnwidth, 0.4\columnwidth) node {a)};
   \end{tikzpicture}
   \begin{tikzpicture}
     \draw (0, 0) node [inner sep=0] {\includegraphics[width=0.95\columnwidth]{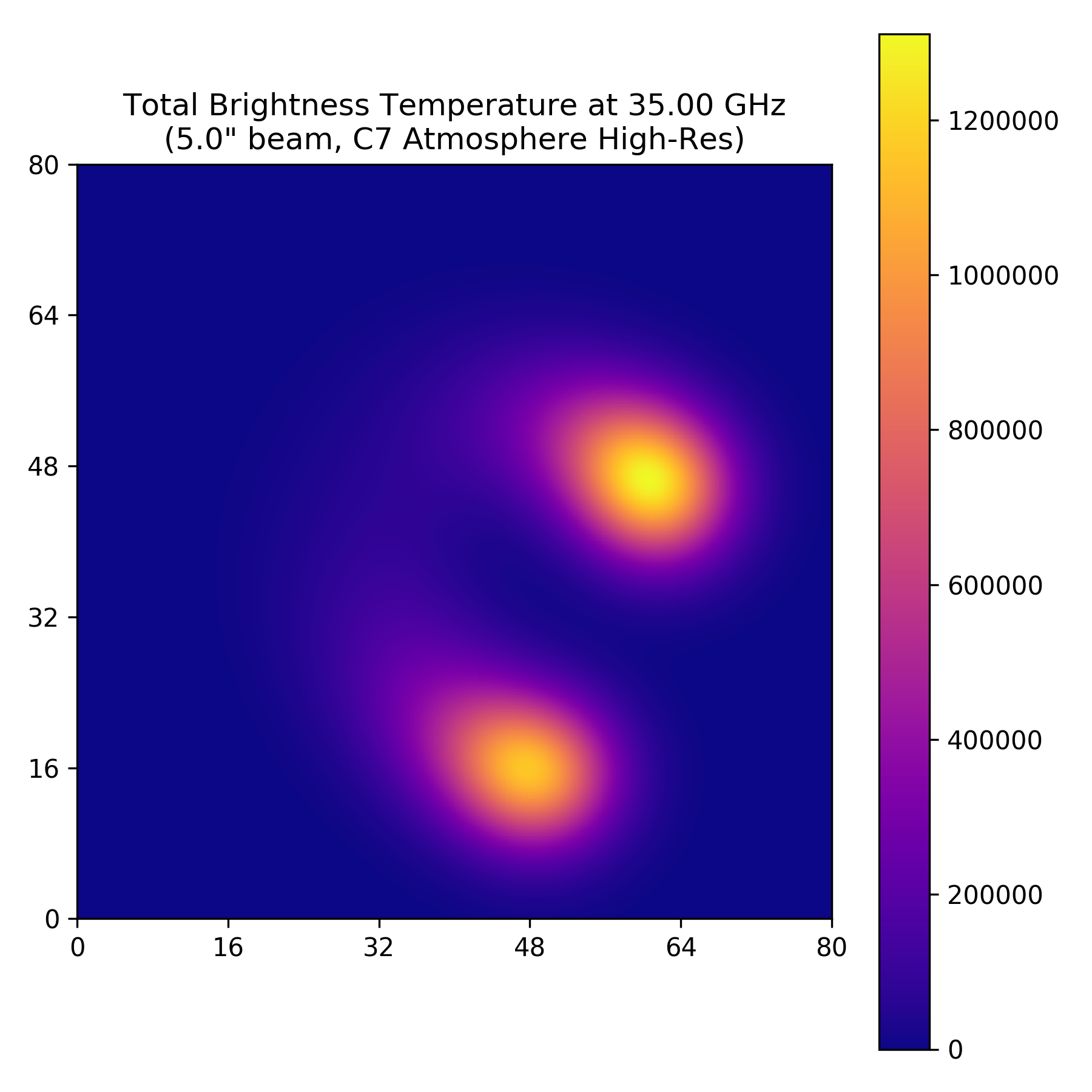}};
     \draw (-0.4\columnwidth, 0.4\columnwidth) node {b)};
   \end{tikzpicture}
  \caption{35 GHz emission map convolved with a 5" beam for the low-- (a)) and high--resolution (b)) simulations}
      \label{Fig:TbConv}
\end{figure*}

The \textit{Thyr} tool, its initial validation, and example simulations have 
now been presented.
The high-resolution simulation presented here is computed at much higher
resolution than is available with modern radio observatories.
This fine structure remains nonetheless important due to the contribution 
of the radiation produced in these regions for the total beam power.
This can be seen by looking at the image that an observatory such as Nobeyama
would produce from the different models.
These convolved maps are shown in Fig.~\ref{Fig:TbConv}.
The maps shown in these figures appear very similar, but the peak brightness
temperature is different, and the magnitude of this difference only increases at higher
frequencies, as demonstrated by the spectra (e.g. Fig~\ref{Fig:C7Flux}).

Using the spectral imaging techniques of Section~\ref{Sec:ImagingSpectro} we have identified the location from which the hardening of the high frequency emission originates.
Points 3, 4 and 5 from Fig~\ref{Fig:OptThick} lie within a thin ($\sim$300 km) 
layer in the upper chromosphere and transition region.
This corresponds to a layer within the region of 1.8--2.15 Mm on
Fig.~\ref{Fig:C7Atm}.
The field chosen for this simulation was a dipole as this represents the
simplest analytical case and a realistic flaring active region can be 
expected to have stronger magnetic convergence.
In this simulation the average field along the line of sight for each of 
these pixels is $\sim$1.3 kG and they lie within a region of strong 
magnetic convergence.
This produces the total hardening at high frequencies due to the 
superposition of the small-scale hardenings produced by emission from 
small regions with high magnetic field strength in the core of this thin slab.

Given its characteristic spectral signature, there is no need
to spatially resolve this thin layer.
If the emission from the footpoints cannot be resolved separately then a similar spectrum could also be obtained from a loop with different magnetic field strengths in the footpoints (e.g. with a significantly rotated magnetic moment).
To verify if the emission behaves in this way increased spectral resolution
at high frequencies is required.
This can be achieved with today's technology using simultaneous observations
from \textit{Radio Solar Telescope Network} \citep[RSTN, operating up to 15.4~GHz,][]{Guidice1979}, \textit{Nobeyama Radio Polarimeters} \citep[NoRP,][]{Nakajima1985}, \textit{Nobeyama Radio Heliograph} \citep[NoRH,][]{Nakajima1994}
\textit{POlarization Emission of Millimeter Activity at the Sun} \citep[POEMAS, operating at 45 and 90~GHz][]{Valio2013}, ALMA
\citep[currently operating at 100 and 230 ~GHz][]{Wedemeyer2016},
and the Submillimeter Solar Telescope \citep[SST, 212 and 405 GHz][]{Kaufmann2008}. The combination of these instruments could provide valuable information on
the shape of the spectrum at higher frequencies and hence an estimate of the structure of the magnetic field in the atmosphere.

\section{Conclusions}

We have presented three-dimensional solar flare radio emission simulation
tool \textit{Thyr}, verification against
\citet{Klein1984}, and example simulations that demonstrate the importance 
of resolving the lower atmosphere to a much higher resolution than used in 
previous models.
This tool is now available for use under the MIT license and can 
be downloaded at \url{http://github.com/Goobley/Thyr2} \citep{Osborne2018}

\textit{Thyr} has the ability to simulate user-specified regions with
increased accuracy, and we use this to increase the resolution in the lower
corona, chromosphere and photosphere.
By performing simulations with a high resolution lower atmosphere we have
shown that a non-thermal hardening of the spectrum should be expected at
higher frequencies from a dipole loop.
Using a combination of RSTN, POEMAS, and ALMA, the existence of such a
hardening could be investigated.

Recent studies have also argued for the importance of thermal free-free
emission at higher frequencies \citep{Simoes2017, Heinzel2012}.
Whilst this emission is not important in the model we have selected here,
\textit{Thyr} could well be used to investigate the parameters for which it
becomes significant.
The C7 atmosphere was chosen here due to its ubiquity and high-resolution --
there is little reason to perform a high resolution simulation with a low
resolution atmosphere!
It is simple to investigate the influence of other atmospheres using
\textit{Thyr} and the files present in the github repository can serve as a base. For example, \textit{Thyr} can accept snapshots of the flaring atmosphere generated by radiative hydrodynamic simulations, such as RADYN \citep{Carlsson1995,Allred2015} and Flarix \citep{Varady2010,Heinzel2017}.

\textit{Thyr} can also serve as a skeleton for future local thermodynamic
equilibrium radiative transfer codes as it is simple to replace the geometry
and/or use different emission and absorption coefficients.
Three-dimensional modelling is especially useful in cases where the emission
has a high angular dependence, such as the case with gyrosynchrotron here.
We therefore hope that this code can be modified and be of use to the wider
astronomical community.

\section*{Acknowledgements}
 The authors would like to thank Lyndsay Fletcher for helpful comments and suggestions.
 C.M.J.O is grateful for the financial support of the Royal Society of Edinburgh 
 Cormack Undergraduate Vacation Research Scholarship (2016), 
 and the Royal Astronomical Society Undergraduate 
 Summer Bursary (2015) under which this project took shape. 
 C.M.J.O also acknowledges financial support from STFC Doctoral Training Grant ST/R504750/1.
 P.J.A.S. acknowledges support from the University of Glasgow's Lord Kelvin Adam Smith Leadership Fellowship. This work builds upon the results obtained from projects funded by FAPESP grants 03/03406-6, 04/14248-5, 08/09339-2 and 2009/18386-7.
 We acknowledge the use of colour-blind safe and print-friendly colour tables 
 by Paul Tol (\url{https://personal.sron.nl/~pault/}). The authors would also like to thank the reviewer for detailed comments and suggestions for improvement.


\bibliographystyle{mnras}
\bibliography{Refs}

\bsp	
\label{lastpage}
\end{document}